\documentclass[%
 aip,
 amsmath,amssymb,
 reprint,%
]{revtex4-1}

\usepackage{graphicx}
\usepackage{dcolumn}
\usepackage{bm}

\usepackage[utf8]{inputenc}
\usepackage[T1]{fontenc}
\usepackage{mathptmx}
\usepackage{etoolbox}

\makeatletter
\def\@email#1#2{%
 \endgroup
 \patchcmd{\titleblock@produce}
  {\frontmatter@RRAPformat}
  {\frontmatter@RRAPformat{\produce@RRAP{*#1\href{mailto:#2}{#2}}}\frontmatter@RRAPformat}
  {}{}
}%
\makeatother
\begin{document}

\preprint{AIP/123-QED}

\title[]{Plasma response to resonant magnetic perturbations near rotation zero-crossing in low torque plasmas}
\author{Pengcheng Xie}
\affiliation{Institute of Plasma Physics, CAS, PO Box 1126, Hefei 230031, China}
\affiliation{University of Science and Technology of China, Hefei 230026, China}
\author{Youwen Sun}\thanks{Author to whom any correspondence should be addressed}
\affiliation{Institute of Plasma Physics, CAS, PO Box 1126, Hefei 230031, China}
\email{ywsun@ipp.ac.cn}
\author{Yueqiang Liu}
\affiliation{General Atomics, PO Box 85608, San Diego, CA 92186-5608, USA}
\author{Shuai Gu}
\affiliation{General Atomics, PO Box 85608, San Diego, CA 92186-5608, USA}
\author{Qun Ma}
\affiliation{Institute of Plasma Physics, CAS, PO Box 1126, Hefei 230031, China}
\affiliation{University of Science and Technology of China, Hefei 230026, China}
\author{Cheng Ye}
\affiliation{Institute of Plasma Physics, CAS, PO Box 1126, Hefei 230031, China}
\affiliation{University of Science and Technology of China, Hefei 230026, China}
\author{Xuemin Wu}
\affiliation{Institute of Plasma Physics, CAS, PO Box 1126, Hefei 230031, China}
\affiliation{University of Science and Technology of China, Hefei 230026, China}
\author{Hui Sheng}
\affiliation{University of Science and Technology of China, Hefei 230026, China}
\author{The EAST Team}
\affiliation{Institute of Plasma Physics, CAS, PO Box 1126, Hefei 230031, China}

\date{June, 2021}

\begin{abstract}
  Plasma response to resonant magnetic perturbations (RMPs) near the pedestal top is crucial for accessing edge localized modes (ELMs) suppression in tokamaks.
  Since radial location of rotation zero-crossing plays a key role in determining the threshold for field penetration of RMP, plasma response may be different in low input torque plasmas. 
  In this work, the linear MHD code MARS-F is applied to reveal the dependence of plasma response to RMP on rotation zero-crossing by a scan of rotation profiles based on an EAST equilibrium.
  It is shown that the plasma response is enhanced when zero-crossing occurs near rational surfaces.
  The dependence of plasma response on the location of rotation zero-crossing is well fitted by a double Gaussian, indicating two effects in this enhancement.
  One is induced by rotation screening effect shown as a wide base (with a width around $10-20~\mathrm{krad/s}$), and the other is related to resistive singular layer effect characterized by a localized peak (with a width around $3-4~\mathrm{krad/s}$).
  The peak of each resonant harmonic in plasma response appears always at rotation zero-crossing.
  The width of the peak scales with the resistive singular layer width. 
  The plasma displacement suggests the response is tearing like when zero-crossing is within the singular layer, while it is kink like when zero-crossing is far from the layer.
  The enhancement of magnetic islands width at the peak is only around a factor of two, when the absolute value of local rotation is not larger than $10-20~\mathrm{krad/s}$.
  It is further confirmed in a modeling of plasma response in an EAST ELM suppression discharge.
  Though there is a zero-crossing in $E\times B$ rotation but not in electron perpendicular rotation, no significant difference in plasma response is obtained using these two rotation profiles.
  This suggests that the rotation near pedestal top should not be far away from zero but may not be necessary to have zero-crossing for accessing ELM suppression.

  \noindent{\it Keywords}: plasma response, RMP, rotation, ELM suppression, resistive singular layer, low torque
\end{abstract}

\maketitle

\section{Introduction}

Type I edge localized modes (ELMs) occurring in H mode is expected to cause enormous transient particle and heat loads on plasma facing components like divertors in ITER \cite{evans2015resonant,loarte2003characteristics}.
ELM suppression by resonant magnetic perturbation (RMP) has been obtained widely in DIII-D \cite{evans2005suppression}, KSTAR\cite{jeon2012suppression}, EAST \cite{sun2016nonlinear} and AUG \cite{suttrop2018experimental}, besides mitigation obtained in JET \cite{liang2007active} and MAST \cite{kirk2013understanding}.
Linear models revealed the optimal spectrum of RMP for ELM control of which the figure of merit is edge resonant component of plasma response or plasma surface displacement near the $X$-points \cite{liu2010full,paz2015observation,liu2016elm,yang2016modelling,sun2016edge,park20183d}.
Using the experiments in high $n$ RMP and high $q_{95}$, a better criterion is found, i.e. the resonant component near the pedestal top \cite{gu2019new}, which is important for producing magnetic islands and stochastic region, prohibiting the increase of pedestal height to trigger ELM \cite{snyder2012eped,wade2015advances}. 

The optimal spectrum is not sufficient for accessing ELM suppression.
Chirikov parameter $>1$ was proposed as the necessary of ELM suppression that overlap of magnetic islands at the edge offers a channel for particle transport, though it was based on vacuum calculation of RMP and didn't consider plasma response \cite{evans2004suppression,fenstermacher2008effect}.
The transition from ELM mitigation to full suppression is a nonlinear process and related to penetration of magnetic perturbations at pedestal top \cite{sun2016nonlinear,nazikian2015pedestal,snyder2012eped,hu2020wide}, which is tightly correlated with plasma rotation, viscosity and resistivity \cite{fitzpatrick1998bifurcated,waelbroeck2012role}.
The experiment in L-mode discharges in TEXTOR shows that the strength of perturbation field for a locked tearing mode reaches a minimum when its frequency equals electron perpendicular rotation ($\omega_{\perp e}$), i.e. the sum of diamagnetic drift frequency ($\omega_{* e}$) and $E\times B$ rotation frequency ($\omega_E$) \cite{koslowski2006dependence}.
Following theoretical studies, including kinetic, linear and non-linear two-fluid models, show a lowest penetration threshold happens when $\omega_{\perp e}$ is zero \cite{heyn2006kinetic,kikuchi2006forced,yu2008numerical,becoulet2012screening}.
It is revealed further that field penetration tends to happen when zero-crossing of $\omega_{\perp e}$ crosses a rational surface \cite{waelbroeck2012role} and that co-alignment of the zero-crossing with a rational surface at pedestal top would enhance resonant harmonic in plasma response resulting in easy field penetration \cite{wade2015advances,nazikian2015pedestal}.
Therefore, zero-crossing of $\omega_{\perp e}$ at pedestal top might be favorable for field penetration and then ELM suppression.

In counter neutral beam injection (NBI) or radio frequency (RF) heating plasma, zero-crossing of $\omega_{\perp e}$ would disappear as toroidal rotation reduces, that might be unfavorable for ELM suppression.
The DIII-D experiment provides an example that ELM suppression couldn't be obtained in low co-$I_p$ torque injection, where the zero-crossing of $\omega_{\perp e}$ shifts inward from pedestal top when toroidal rotation decreases \cite{moyer2017validation}.
A simulation work shows order of magnitude distinct effect for plasma response when rotation is zero or non-zero at a rational surface by scanning of rotation zero-crossing \cite{lyons2017effect}.
In contrast, ELM suppression is obtained in low torque plasma with radio-frequency dominant heating in EAST, with no zero-crossing of $\omega_{\perp e}$ (several $\mathrm{krad/s}$) at pedestal top \cite{sun2016nonlinear}.
Recent work in DIII-D also observed zero-crossing of $\omega_{\perp e}$ absent or deviating from a rational surface during ELM suppression \cite{paz2019effect}.
These results raise the question what role the zero-crossing of $\omega_{\perp e}$ plays and whether it is necessary to hold it for ELM suppression.
Adopting similar technique in Ref.\cite{lyons2017effect}, a systematic modeling based on an EAST low input torque equilibrium has been carried out studying the dependence of plasma response on rotation zero-crossing with linear resistive MHD model MARS-F by a scan of rotation zero-crossing \cite{liu2000feedback,turnbull2013comparisons}.

The paper is structured as follows.
In section \ref{sec:models}, the EAST equilibrium and rotation profile model employed in this study are described. 
Section \ref{sec:results} gives generic features of the dependence of resonant harmonic in plasma response on rotation zero-crossing and the effect of plasma density, resistivity and rotation shear.
Section \ref{sec:application} applies the modeling results for understanding ELM suppression in low torque plasma through comparison of modeling results using rotation profiles with or without zero-crossing, followed by a summary in section \ref{sec:conclusion}.

\section{Numerical models and equilibrium setup}
\label{sec:models}
In this section, the model of rotation profiles with assigned zero-crossing and plasma equilibrium are introduced.
MARS-F is a single fluid linear model including plasma resistivity and rotation, which has been well validated by many experiments \cite{lanctot2010validation,wang2015three}.
The linear MHD equations solved in the MARS-F code can be written as \cite{liu2011modelling}
\begin{eqnarray}
  &&i\left(\Omega_\mathrm{RMP}+n \Omega\right) \xi = v +( \xi \cdot \nabla \Omega) R \hat{\phi},    \label{eqn:mars1}                                                                                                                                                                                                                             \\
  &&i \rho\left(\Omega_\mathrm{RMP}+n \Omega\right) v =-\nabla p+ j \times B + J \times b                          -\rho[2 \Omega \hat{ Z } \times v      \nonumber\\
    &&\qquad+( v \cdot \nabla \Omega) R \hat{\phi}]-\rho \kappa_{\|}\left|k_{\|} v_{ th , i }\right|\left[ v +( \xi \cdot \nabla) V _{0}\right]_{\|}, \label{eqn:mars2} \\
  &&i\left(\Omega_\mathrm{RMP}+n \Omega\right) b=\nabla \times( v \times B )+( b \cdot \nabla \Omega) R \hat{\phi}  -\nabla \times(\eta j ), \label{eqn:mars3}                                                                                                                                                                                    \\
  &&i\left(\Omega_\mathrm{RMP}+n \Omega\right) p=- v \cdot \nabla P-\Gamma P \nabla \cdot v, \label{eqn:mars4}                                                                                                                                                                                                                                    \\
  &&j=\nabla \times b, \label{eqn:mars5}
\end{eqnarray}
where $V_0=R\Omega\hat{\phi}$, and the variables $\eta$, $\xi$, $v$, $b$, $j$, $p$, $\Omega$ represent the plasma resistivity, plasma displacement, perturbed velocity, magnetic field, current, pressure, and toroidal angular frequency, respectively.

\begin{figure}[htbp]
  \centering
  \includegraphics[width=0.48\textwidth]{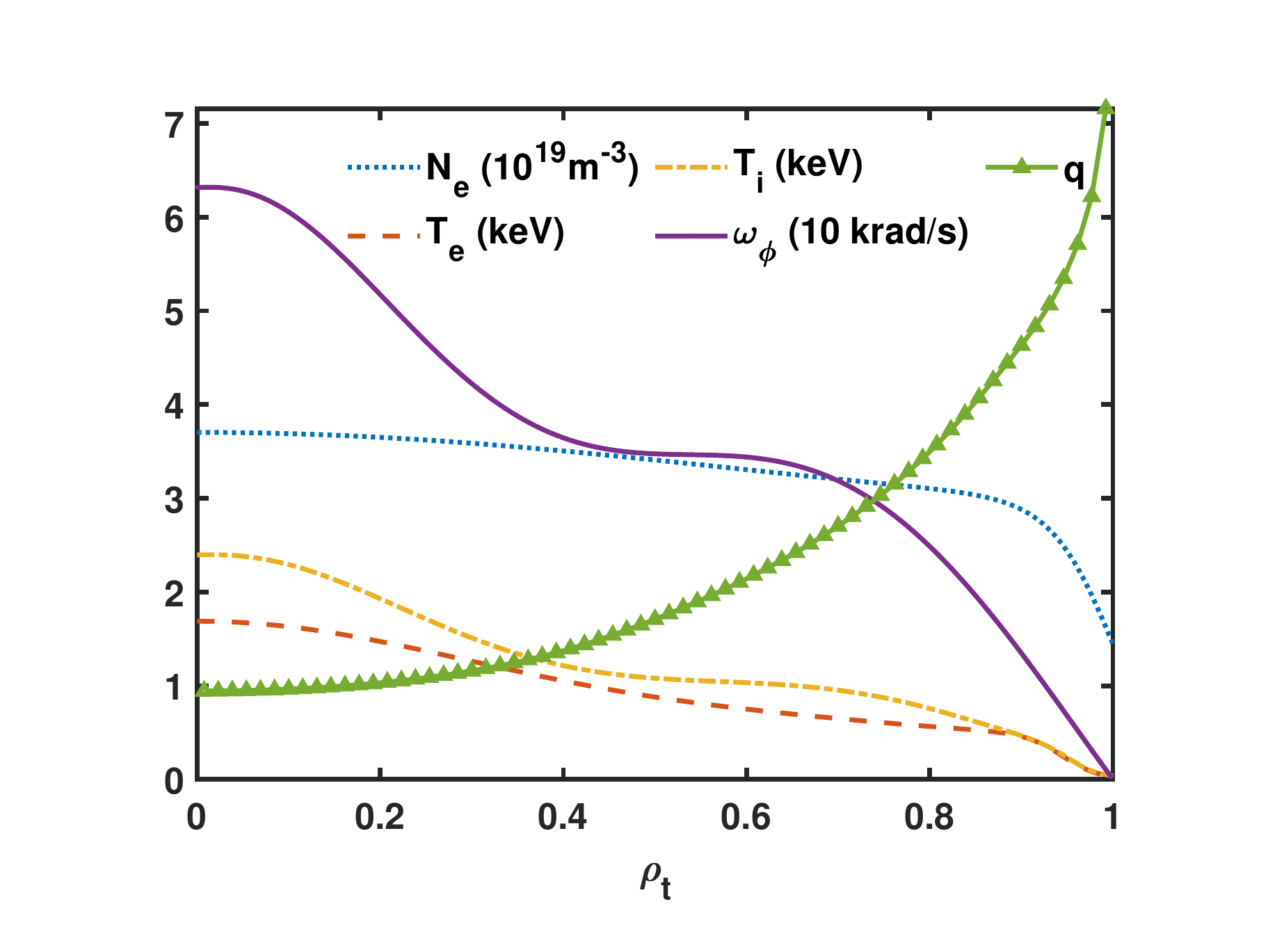}
  \caption{Radial profiles of plasma density ($N_e$, dotted line), temperature of electron and ion ($T_e$, dashed line; $T_i$, dashed-dotted line), plasma toroidal rotation ($\omega_{\phi}$, solid line), safety factor ($q$, triangle line) in the EAST discharge 52340 at 3150 ms.}
  \label{mfig:prof52340pre}
\end{figure}

EAST, as a superconducting tokamak, possesses the features such as low torque and ITER-like divertors \cite{wan2019recent,wan2020new}.
An example plasma equilibrium of EAST discharge 52340 at 3150 ms are reconstructed with K-EFIT \cite{sun2016edge} is used as a reference.
Figure \ref{mfig:prof52340pre} shows the radial profiles of plasma density ($N_e$, dotted line), temperature of ion ($T_i$, dashed-dotted line) and electron ($T_e$, dashed line), toroidal rotation ($\omega_{\phi}$, solid line) and safety factor ($q$, triangle line).
Here, $\rho_t=\sqrt{\psi_t}$ is the normalized radius, and $\psi_t$ is the normalized toroidal flux.
$q=5$ rational surface is located at $\rho_t=0.926$.

In this reference case, Alfv\'en time $\tau_A$ at the magnetic axis is $3\times 10^{-7}~\mathrm{s}$ and the resistive diffusion time at the magnetic axis $\tau_R=\mu_0a^2/\eta\approx10~\mathrm{s}$. 
The natural mode frequency in two fluids model can be written as $\omega_{\mathrm{MHD}}=\omega_{e\perp}=\omega_E+\omega_{*e}$, while it reduces to $\omega_E$ in single fluid model.
Here $\omega_E$ is the $E\times B$ frequency and $\omega_{*e}$ is the electron diamagnetic frequency.
Previous studies \cite{yu2008numerical,liu2014modelling,cole2006drift} showed that there is no significant difference between the results in field penetration threshold obtained from the single fluid model with $\Omega=\omega_E+\omega_{*e}$ and the two fluids model.
Therefore, we just choose different flow profiles of $\Omega$ to study different effects in the following modeling, which was applied in previous studies \cite{liu2017comparative}.

\begin{figure}[htbp]
  \centering
  \includegraphics[width=0.48\textwidth]{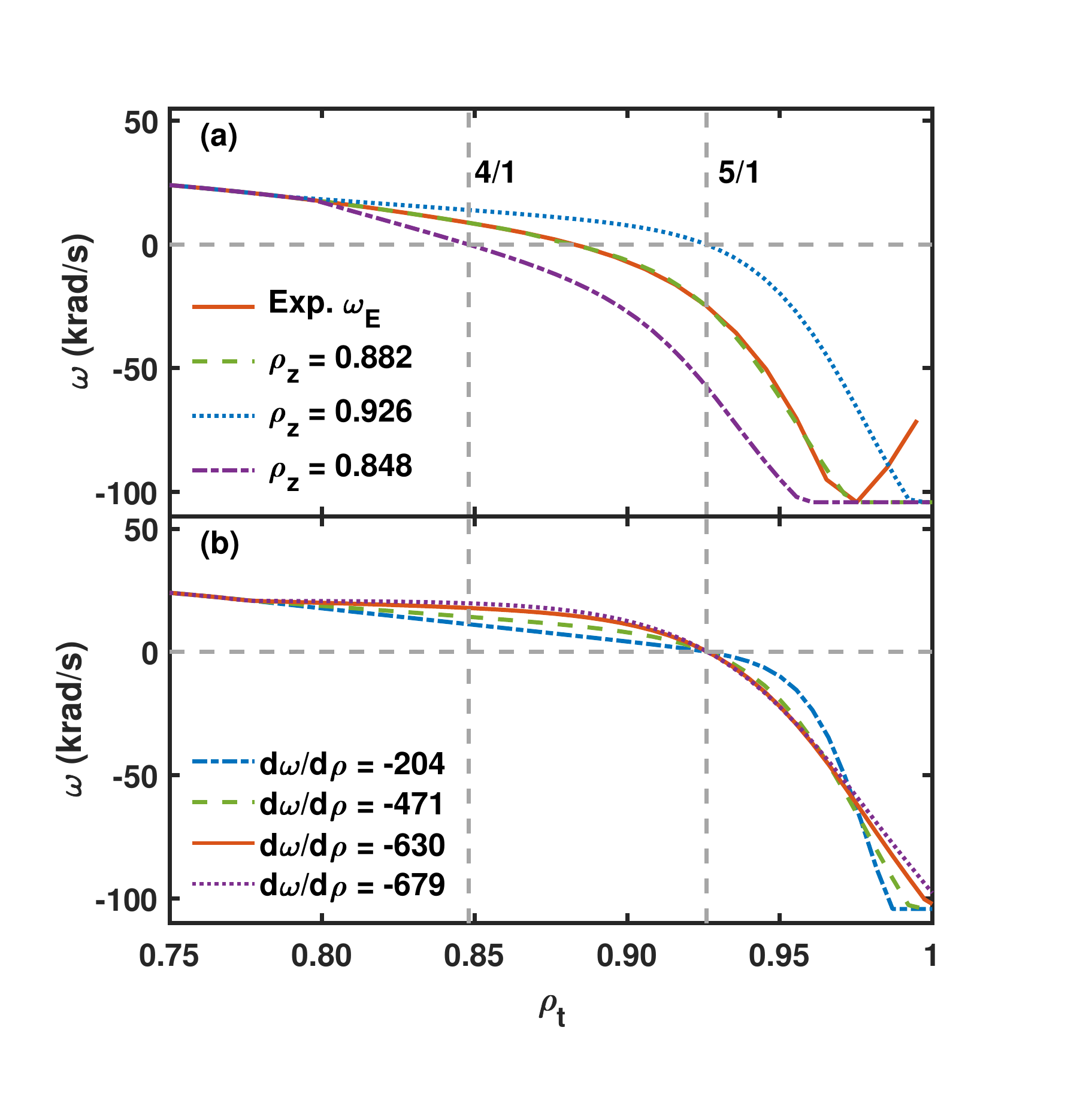}
  \caption[resonant component with zero-crossing of $\Omega_{E\times B}$]
  {Examples of rotation profiles used in the following modeling.
    (a) Rotation profiles with assigned location of zero-crossing, $\rho_z=0.882$ (dashed line), $0.848$ (dashed-dotted line) and $0.926$ (dotted line).
    Among them, the profile with $\rho_z=0.882$ coincides the experimental $\omega_E$ profile (solid line).
    Vertical dashed lines point out the location of $q=4,~5$ rational surfaces respectively.
    (b) Rotation profiles with different shear at zero-crossing, as $\mathrm{d}\omega/\mathrm{d}\rho=-204~\mathrm{krad/s}$ (dashed-dotted line), $-471~\mathrm{krad/s}$ (dashed line, the shear in (a)), $-630~\mathrm{krad/s}$ (solid line), $-679~\mathrm{krad/s}$ (dotted line) with $\rho_z=0.926$.}
  \label{mfig:rotationprofile}
\end{figure}

The dependence of plasma response on rotation zero-crossing is obtained by scanning of rotation profiles for MARS-F modeling \cite{liu2017comparative}.
Hyperbolic tangent is employed to construct rotation profiles with assigned zero-crossing applying similar method in Ref. \cite{lyons2017effect}.
Figure \ref{mfig:rotationprofile} shows the examples of two kinds of scans of rotation, e.g. location of zero-crossing and rotation shear at zero-crossing, used in the modeling.
Figure \ref{mfig:rotationprofile}a shows one reference profile according to experimental $\omega_{E}$ (solid line) and modeling profiles with different zero-crossing locations ($\rho_z$) as $\rho_z=0.882,~0.848,~0.926$ (dashed line, dashed-dotted line, dotted line respectively).
Rotation profiles in Fig.\ref{mfig:rotationprofile}b have different shear at $q=5$ surface as $\mathrm{d}\omega/\mathrm{d}\rho=-204,-471,-630,-679 ~\mathrm{krad/s}$ (dashed-dotted line, dashed line, solid line and dotted line, separately) with $\rho_z=0.926$.
Rotation shear of the reference profile around $q=5$ surface in Fig.\ref{mfig:rotationprofile}a is $-471~\mathrm{krad/s}$.
Locations of rational surfaces $q=4,~5$ are indicated by vertical dashed lines.
A fine scan of rotation zero-crossing is performed to resolve the detailed dependence of plasma response as zero-crossing passing through the rational surface.
It should be noted that location of zero-crossing has to avoid mesh nodes around rational surfaces for eliminating numerical errors in the modeling.
Otherwise, artificial numerical peaks may appear in the results, because of the singularity in the equations.

RMP configuration is set as only upper coils in simulation for simplicity (detailed information of EAST RMP coils in Ref. \cite{sun2015modeling}), and current of RMP coils is $2.5~\mathrm{kA}$ with 4 turns and toroidal mode number $n=1$.
Since results at different rational surfaces are similar, only the plasma response around $q=5$ surface is discussed in the following.

\section{Modeling the dependence of plasma response on zero-crossing}
\label{sec:results}
The dependence of plasma response on rotation zero-crossing shows two effects with different scales and is well fitted by a double Gaussian function consisting with a localized peak and a wide base.
Influence of plasma density, resistivity and rotation shear on the dependence suggests that the two Gaussian models represent rotation screening effect and resistivity singular layer effect, separately.
Importantly, it is found that the resonant harmonic in plasma response mainly depends on the absolute value of local rotation.

\subsection{Main characteristics of resonant response}
\begin{figure}[htbp]
  \centering
  \includegraphics[width=0.48\textwidth]{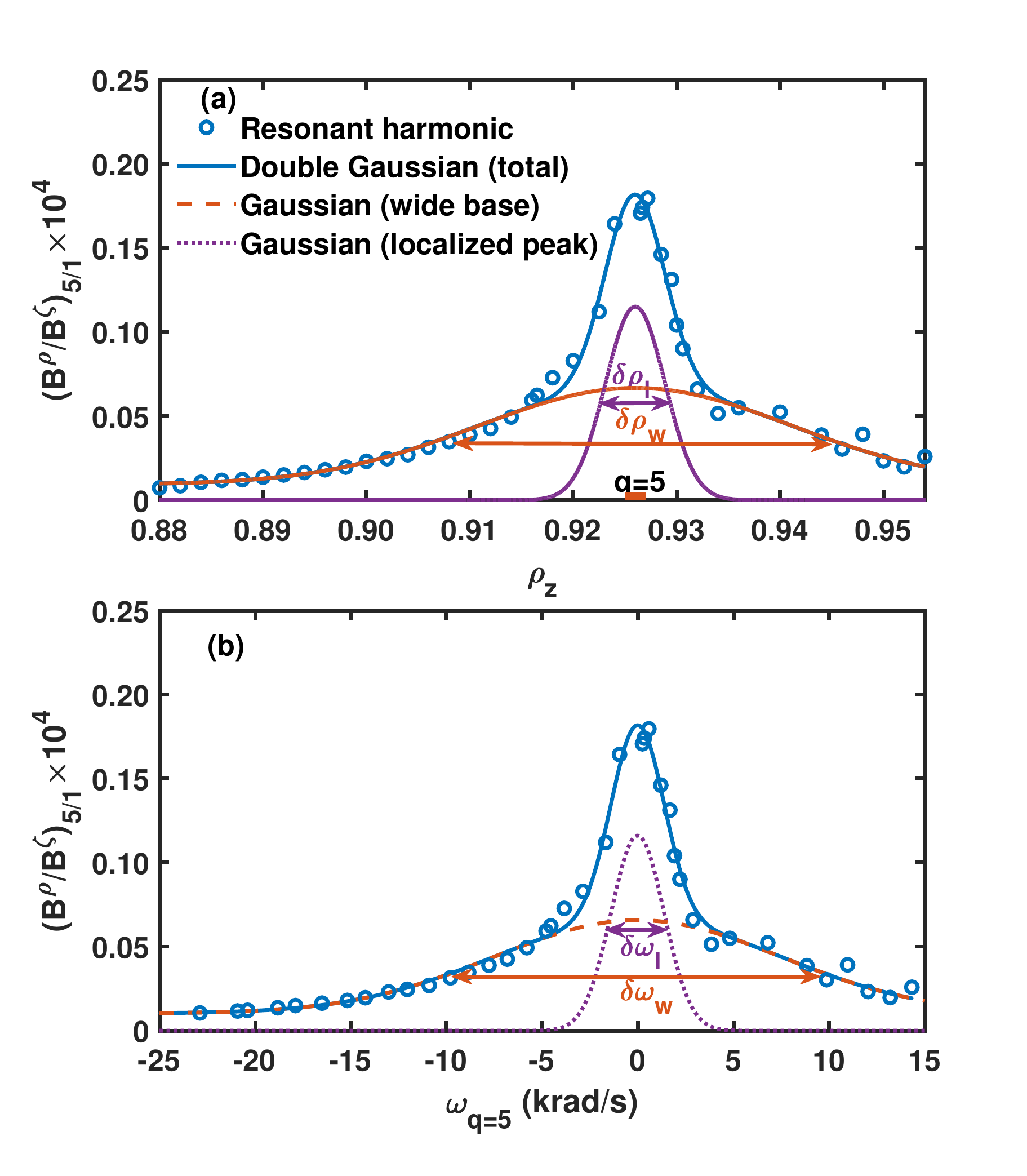}
  \caption{(a) Dependence of resonant $m/n=5/1$ Fourier component of magnetic perturbations with plasma response field at $q=5$ surface (circles) on rotation zero-crossing location, and its fitted curve with a double Gaussian function defined in Eq.\ref{eq:doubleGaussian} (solid line).
    Full widths at half maximum (FWHM) of the two scales Gaussians are defined as $\delta\rho_w$ (wide base: dashed line, $\sim3.58\%$ of $\rho_t$), $\delta\rho_l$ (localized peak: dotted line, $\sim0.68\%$ of $\rho_t$) respectively.
    Amplitude of them is defined as $\delta B_w$ ($\sim0.057$) and $\delta B_l$ ($\sim0.115$) respectively.
    (b) Dependence of the resonant harmonic (circles) and its fitted curve with the double Gaussian (solid line) on rotation velocity at $q=5$ surface.
    FWHM of its two components (wide base: dashed line, localized peak: dotted line) are defined as $\delta\omega_w$ (wide base, $\sim17.7~\mathrm{krad/s}$) and $\delta\omega_l$ (localized peak, $\sim3.26~\mathrm{krad/s}$)) respectively.
    Amplitudes of them are nearly $\delta B_w$ and $\delta B_l$.}
  \label{mfig:B-Te-fit2-pre2}
\end{figure}

Main characteristics of the dependence of plasma resonant response on rotation zero-crossing are investigated firstly.
Here, the plasma response to magnetic perturbations, denoted as $\delta B$, is decomposed into Fourier serises in flux coordinates, and $B^\zeta$ is the magnetic field in toroidal direction.
\begin{equation}
  B^\rho/B^\zeta=\sum_{mn} (B^\rho/B^\zeta)_{mn}\mathrm{e}^{i(m\theta-n\zeta)}
\end{equation}
A Fourier component is called resonant component if it is at $q=m/n$ rational surface.
Figure \ref{mfig:B-Te-fit2-pre2}a shows the dependence of $m/n=5/1$ Fourier component with plasma response at $q=5$ rational surface (circles) on rotation zero-crossing in $\rho_t$.
It is in general consistent with previous understanding that plasma response is enhanced when the rotation at the rational surface reduces to zero.
However, the details of the profile show interesting dependences.
The dependence has a wide base and a localized peak obviously, implying two different scales distributions.
Therefore, we use a double Gaussian function to fit the dependence (solid line), which is defined as
\begin{equation}
  \delta B_{mn}=\delta B_l\cdot\mathrm{e}^{-\frac{1}{4\ln2}(\frac{\rho_z-\rho_0}{\delta\rho_l})^2}+\delta B_w\cdot\mathrm{e}^{-\frac{1}{4\ln2}(\frac{\rho_z-\rho_0}{\delta\rho_w})^2}+c
  \label{eq:doubleGaussian}
\end{equation}
$\delta\rho_l$ and $\delta\rho_w$ are full width at half maximum (FWHM) of the two Gaussians, separately.
The symmetry axes $\rho_0$ of the double Gaussian are set the same, because not only they are little different when set as two independent quantities, but also the two effects characterized by the double Gaussian ought to center with the rational surface in general cases, which is shown below.
Fitting shows that the symmetry axis is located at $q=5$ rational surface exactly ($\rho_0=0.926$), which means the resonant harmonic reaches maximum when rotation is zero at the rational surface.
The two constituent Gaussians possess strikingly different scales.
The width of base Gaussian (dashed line) $\delta\rho_w$ is around $3.6\%$ of $\rho_t$, while that of the localized peak Gaussian (dotted line) $\delta\rho_l$ is only $0.7\%$ of $\rho_t$.
Contrary to previous study \cite{lyons2017effect}, the enhancement of resonant Fourier component at the peak is only around a factor of 3, showing that plasma response is not so different when rotation zero-crossing is within $\delta\rho_w$.

Figure \ref{mfig:B-Te-fit2-pre2}b shows the dependence of the same resonant harmonic on rotation velocity at $q=5$ surface (replace $\rho_z$ by the angular velocity at $q=5$ surface $\omega_{q=5}$ and $\rho_0$ by symmetry axis $\omega_0$ in Eq.\ref{eq:doubleGaussian}).
The dependence is also well fitted by a double Gaussian function with a symmetric axis at zero rotation.
FWHM of the two constituent Gaussians are around 18 krad/s and 3 krad/s for the wide base ($\delta\omega_w$) and the localized peak ($\delta\omega_l$) respectively.
Distinct scales of the two Gaussians also suggest two effects in the enhancement.
The wide base Gaussian reflects the rotation screening effect \cite{liu2011modelling,becoulet2012screening}, and the localized peak Gaussian represents the influence of resistive singular layer of tearing modes \cite{waelbroeck2012role}.
Although the localized peak Gaussian is larger near the rational surface, the enhancement of driven magnetic island width is less than around a factor of 2 when $\omega_{q=5}$ is within $\delta\omega_{w}$.
Therefore, it may not be necessary for rotation zero-crossing being exactly at a rational surface in the case of field penetration for accessing ELM suppression. 
\subsection{Understanding of the localized peak and wide base}
Towards understanding the physical meanings of localized peak and wide base profiles in plasma response, dependence of plasma response on rotation zero-crossing is further investigated by scaling plasma density, resistivity and rotation shear.

\begin{figure}[htbp]
  \centering
  \includegraphics[width=0.48\textwidth]{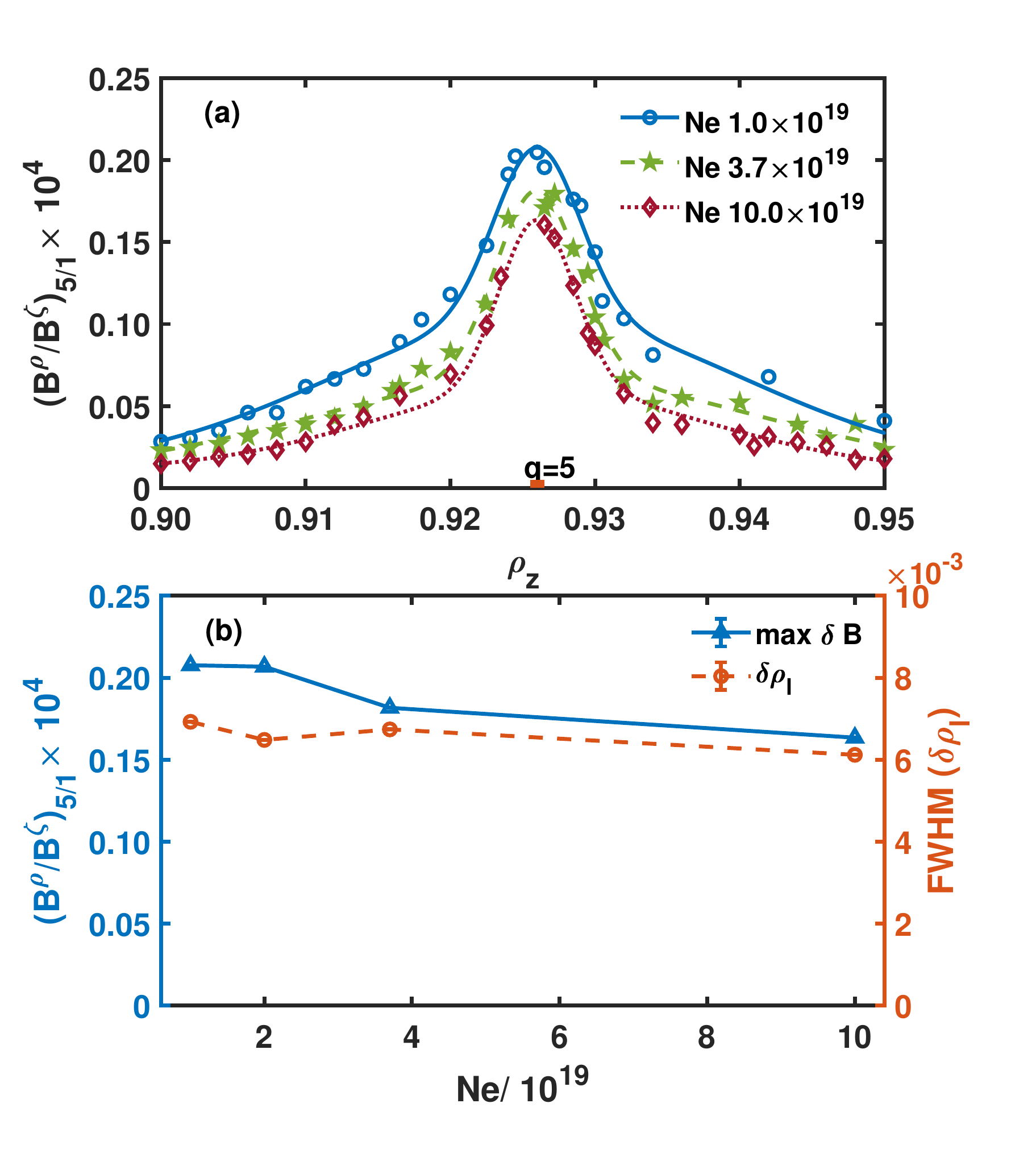}
  \caption{(a) Dependence of resonant $m/n=5/1$ Fourier component of magnetic perturbations with plasma response at $q=5$ surface on rotation zero-crossing location, and its fitted curve with a double Gaussian function, for different plasma densities $N_e=1\times10^{19}$ (circles with solid line), $3.7\times10^{19}$ (pentagrams with dashed line) and $1\times10^{20}~\mathrm{m}^{-3}$ (diamonds with dotted line).
    Among them, $N_e=3.7\times 10^{19}~\mathrm{m}^{-3}$ is the reference case shown in Fig.\ref{mfig:prof52340pre}.
    (b) Dependences of resonant peak amplitude (triangles with solid line) and FWHM of localized peak Gaussian $\delta\rho_l$ (circles with dashed line) on plasma density.}
  \label{mfig:B-Ne-fit2-FWHM-3}
\end{figure}

Figue \ref{mfig:B-Ne-fit2-FWHM-3} shows the dependence of resonant $m/n=5/1$ Fourier component on zero-crossing by scaling plasma density with other parameters fixed.
Resonant component and its fitted curve for plasma density at the core $N_e=1.0\times10^{19},~3.7\times10^{19},~10.0\times10^{19}~\mathrm{m}^{-3}$ are represented as circles with solid line, pentagrams with dashed line and diamonds with dotted line respectively.
The symmetry axis of double Gaussian isn't affected by change of density, remaining at the rational surface.
The resonant harmonic becomes weaker as plasma density arises, which can be seen clearly in Fig.\ref{mfig:B-Ne-fit2-FWHM-3}b (triangles with solid line).
It is also shown in Fig.\ref{mfig:B-Ne-fit2-FWHM-3}b the dependence of width of localized peak Gaussian $\delta\rho_l$ (circles with dashed line) on plasma density, which doesn't show obvious correlation.
This scaling of density suggests that the enhancement of resonant harmonic has little relationship with the Alfv\'en resonance \cite{liu2012continuum}. 

\begin{figure}[htbp]
  \centering
  \includegraphics[width=0.48\textwidth]{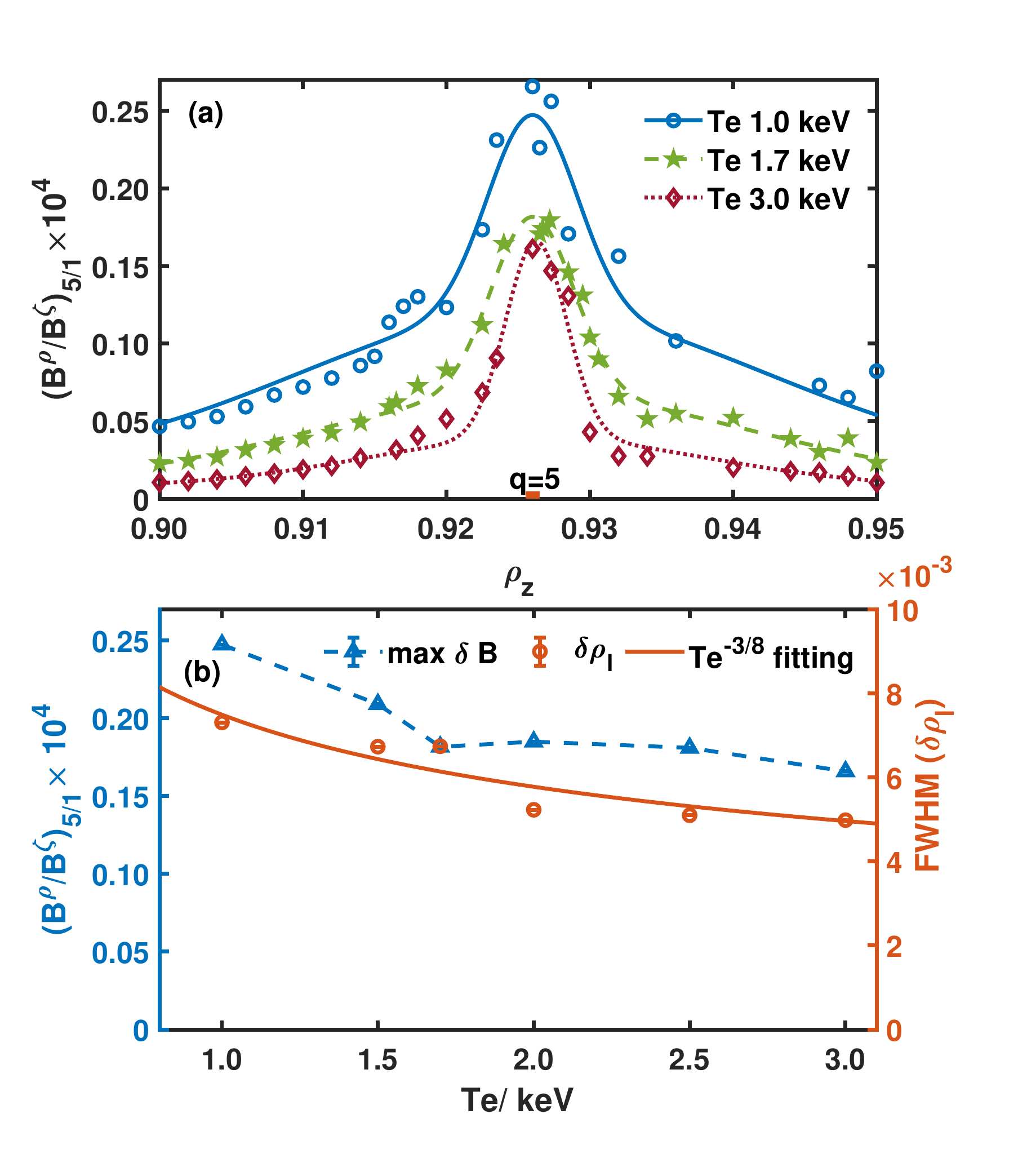}
  \caption{(a) Dependence of resonant $m/n=5/1$ Fourier component of magnetic perturbations with plasma response at $q=5$ surface on rotation zero-crossing, and its fitted curve with a double Gaussian function for different plasma temperature $T_e=1~\mathrm{keV}$ (circles with solid line), $1.7~\mathrm{keV}$ (pentagrams with dashed line) and $3~\mathrm{keV}$ (diamonds with dotted line).
  Among them, $T_e=1.7~\mathrm{keV}$ is the reference one shown in Fig.\ref{mfig:prof52340pre}.
  (b) Dependences of resonant peak amplitude (triangles with dashed line) and FWHM of localized peak Gaussian $\delta\rho_l$ (circles) on plasma temperature.
  The latter is fitted by $a\cdot T_e^{-8/3}$ (solid line).}
  \label{mfig:B-Te-fit2-FWHM-3}
\end{figure}
For the case of scaling on electron temperature, with others parameters fixed, the dependence of resonant harmonic ($m/n=5/1$) on rotation zero-crossing is shown as figure \ref{mfig:B-Te-fit2-FWHM-3}.
Resonant component and its fitted curve for $T_e=1.0,~1.7,~3.0~\mathrm{keV}$ are represented as circles with solid line, pentagrams with dashed line and diamonds with dotted line, separately.
The amplitude of resonant harmonic increases with decreasing plasma temperature (triangles with dashed line).
Since low electron temperature leads to high plasma resistivity, the increased resistivity reduces plasma screening of perturbation field \cite{waelbroeck2012role}. 
The width of the peak $\delta\rho_l$ (circles) also increases with decreasing electron temperature as shown in Fig.\ref{mfig:B-Te-fit2-FWHM-3}b.
It scales well with the resistive singular layer width as $a\cdot T_e^{-3/8}$ (solid line) as shown in Fig.\ref{mfig:B-Te-fit2-FWHM-3}b \cite{wesson2011tokamaks}, and for $T_e=1.7~\mathrm{keV}$ the estimation of the layer width is around $2.2\times10^{-3}$ in $\rho_t$ at $q=5$ surface.
This suggests that the narrow peak is limited to the resistive singular layer physics.

\begin{figure}[htbp]
  \centering
  \includegraphics[width=0.48\textwidth]{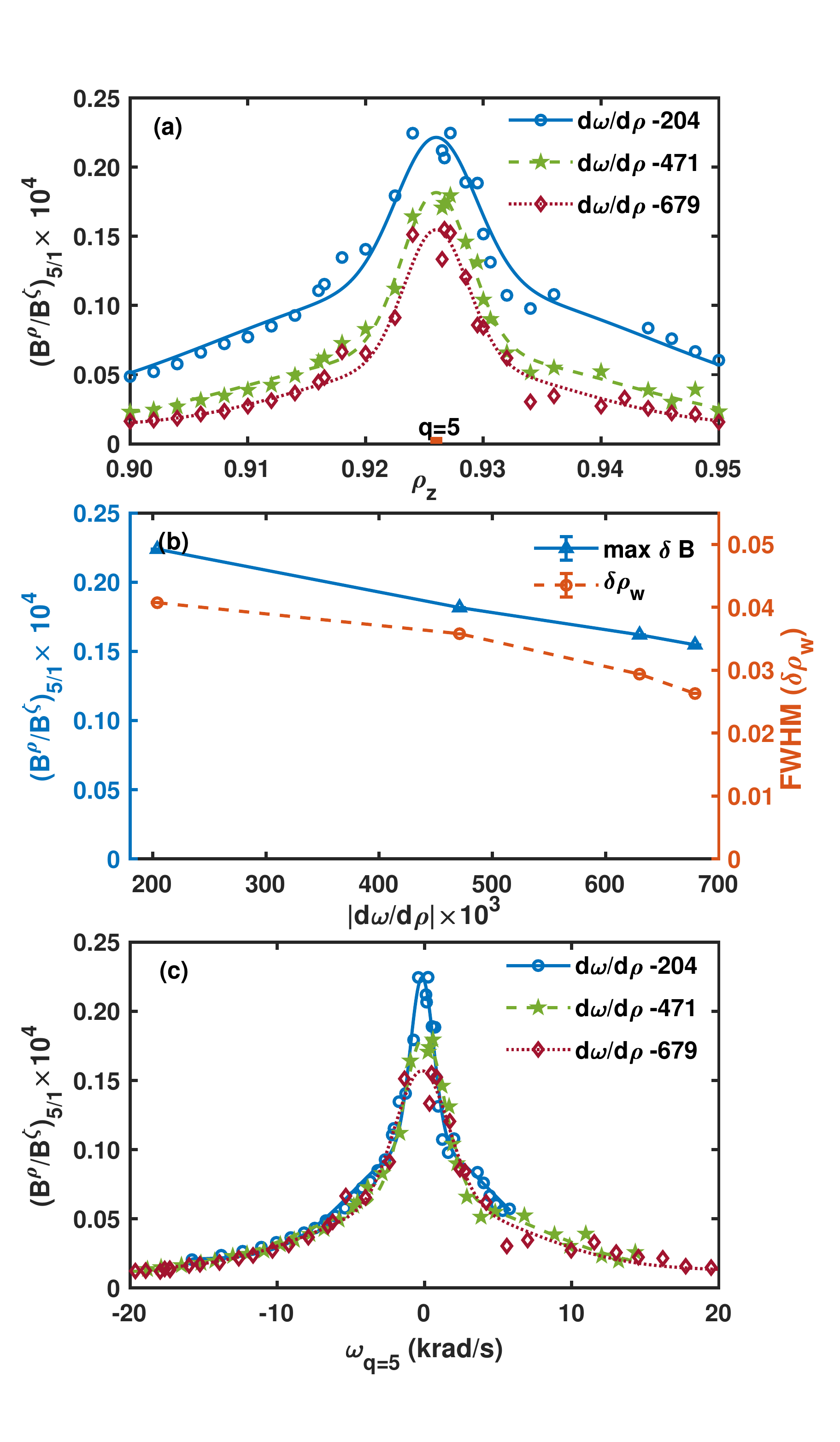}
  \caption{(a) Dependence of resonant $m/n=5/1$ Fourier component of magnetic perturbations with plasma response at $q=5$ surface on rotation zero-crossing, and its fitted curve with a double Gaussian function for different rotation shear at zero-crossing $\mathrm{d}\omega/\mathrm{d}\rho=-204~\mathrm{krad/s}$ (circles with solid line), $-471~\mathrm{krad/s}$ (pentagrams with dashed line) and $-679~\mathrm{krad/s}$ (diamonds with dotted line).
    (b) Dependences of resonant peak amplitude (triangles with solid line) and FWHM of wide base Gaussian $\delta\rho_w$ (circles with dashed line) on rotation shear.
    (c) Dependence of resonant $m/n=5/1$ Fourier component of magnetic perturbations with plasma response field at $q=5$ surface on rotation angular velocity at the surface, and its fitted curve with a double Gaussian function for different rotation shear with the same denotation in (a).}
  \label{mfig:B-slope-fit2-FWHM}
\end{figure}

Figure \ref{mfig:B-slope-fit2-FWHM} shows the dependence of the resonant harmonic on rotation zero-crossing with different rotation shear at zero-crossing and others parameters fixed.
Resonant component and its fitted curve for $\mathrm{d}\omega/\mathrm{d}\rho=-204~\mathrm{krad/s},~-471~\mathrm{krad/s},~-679~\mathrm{krad/s}$ are represented as circles with solid line, pentagrams with dashed line and diamonds with dotted line respectively.
The symmetric axis also keeps at the rational surface.
Under lower rotation shear, amplitude of the resonant harmonic is larger.
The maximum of resonant harmonic, illustrated as triangles with solid line in Fig.\ref{mfig:B-slope-fit2-FWHM}b, is linearly dependent on rotation shear, which is consistent with the theory that fluctuation could be suppressed by flow shear \cite{hahm1995flow}.
It is clearly shown in Fig.\ref{mfig:B-slope-fit2-FWHM}b that the wide base Gaussian becomes broader with lower shear (dashed line), indicating an enhancement of rotation screening effect.

\begin{figure}[htbp]
  \centering
  \includegraphics[width=0.48\textwidth]{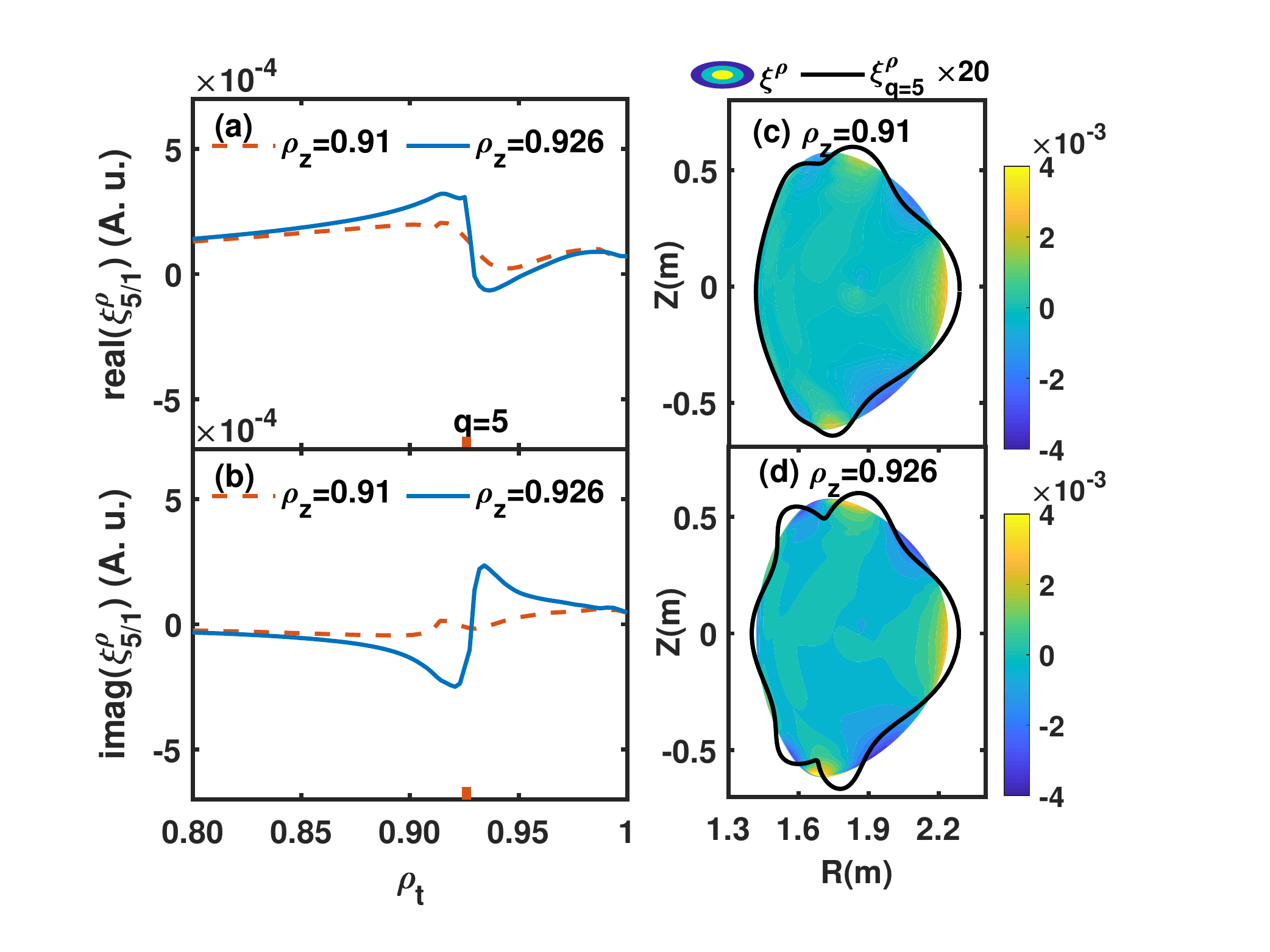}
  \vspace{-5mm}
  \caption{Radial profiles of real part (a) and imaginary part (b) of $m/n=5/1$ Fourier components of plasma radial displacement $\xi^\rho_{m/m}$ induced by $n=1$ magnetic perturbations for $\rho_z=0.926$ (solid line), $0.91$ (dashed line).
    Plasma radial displacement $\xi^\rho$ at $\phi=0^\circ$ poloidal section and a perturbed surface at $q=5$ surface with displacements multiplied by a factor of 20 for $\rho_z=0.91$ (c) and $\rho_z=0.926$ (d).}
  \label{mfig:xi-2d2}
\end{figure}

Although amplitude of the resonant harmonic appears significant distinct for different rotation shear using coordinate of $\rho_z$, it becomes very similar when it is plotted versus rotation velocity at $q=5$ surface as shown in figure \ref{mfig:B-slope-fit2-FWHM}c.
Outside the singular layer around the rational surface, amplitudes of the resonant component for different shears are almost overlapped and isn't affected by rotation shear.
This indicates that absolute value of rotation at the rational surface determines the strength of resonant harmonic.
Here, the differences appearing inside the resonant peak reveals the enhancement of shielding effect by rotation shear via layer physics.
Figure \ref{mfig:xi-2d2} shows radial profiles of real (a) and imaginary (b) parts of Fourier component $m/n=5/1$ of plasma radial displacement, as well as their distribution in a poloidal section (c) and (d).
The displacement around the rational surface is tearing type as rotation zero-crossing is inside the resistive singular layer ($\rho_z=0.926$, solid line), while it is kink type when zero-crossing is outside the layer ($\rho_z=0.91$, dashed line).
The ratio between the displacement at the low field side and high field side is $23$ for $\rho_z=0.91$ and $2.5$ for $\rho_z=0.926$.
These suggest that layer physics dominates in the tearing like response so that plasma resistivity and the rotation shear at the rational surface matter, while the absolute value of rotation determines the kink like response \cite{liu2011modelling}.
Additionally, the coupling effect of $4/1$ or other adjacent Fourier harmonics in plasma response is little, for their low level amplitude or not significant change for different rotation shear.
In conclusion, investigation of rotation shear shows that the absolute value of rotation at the rational surface determines the resonant harmonic in plasma response field and an additional shielding effect of rotation shear via layer physics.

\section{Application of modeling results for understanding ELM suppression in low torque plasma}
\label{sec:application}
Based on the above results, similar plasma response at pedestal top is obtained for modeling results with and without rotation zero-crossing, indicating rotation zero-crossing isn't the essential factor for field penetration.
Besides, predication for measurements in magnetic probes are provided with the linear MHD response model for future work.
\subsection{Plasma response at pedestal top}
\begin{figure}[htbp]
  \centering
  \includegraphics[width=0.48\textwidth]{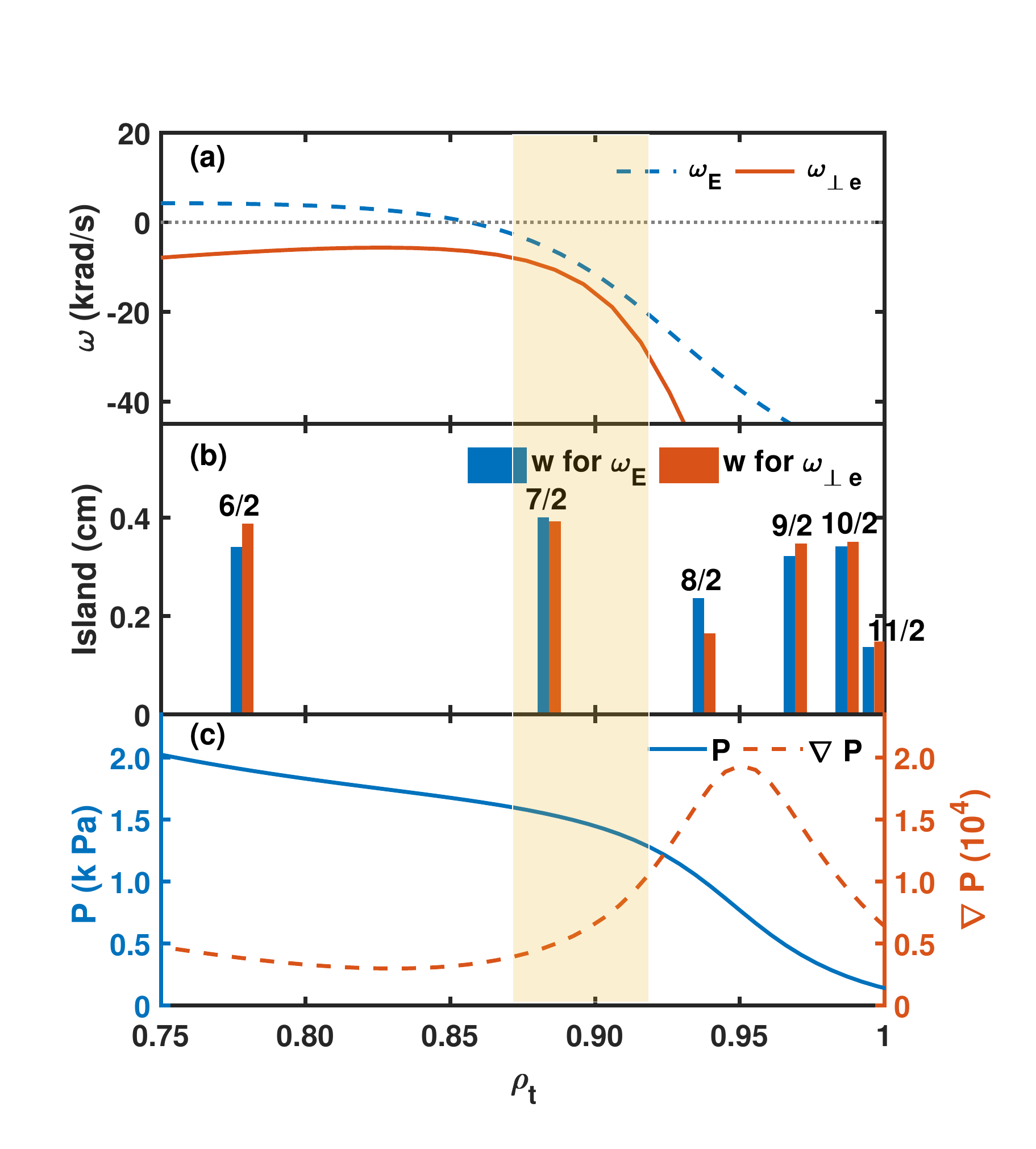}
  \vspace{-5mm}
  \caption{(a) The $\omega_{\perp e}$ (solid line) and $\omega_E$ (dashed line) profile during ELM suppression in EAST discharge 56365. The dotted line indicates $\omega=0$.
    (b) Corresponding magnetic islands width located at rational surfaces to the two rotation profiles.
    (c) Plasma pressure (solid line) and pressure gradient (dashed line) along the radius.
    Shading shows pedestal top area ($\rho_t\approx0.9$).}
  \label{mfig:Brot56366}
\end{figure}
In the following, an analysis of plasma response at pedestal top is presented.
The equilibrium is based on EAST discharge 56365 during RMP-ELM suppression, where $\omega_{\perp e}$ has no zero-crossing, while $\omega_{E}$ has zero-crossing. 
Figure \ref{mfig:Brot56366}a shows the $\omega_{\perp e}$ (solid line) and $\omega_E$ (dashed line) profiles, and (b) shows the width of magnetic islands induced by $n=1$ RMP in these two cases.
In fact, the width of magnetic island at $q=7/2$ rational surface near the pedestal top is not significant different. 
Using $\omega_E$ induces wider magnetic island at $q=8/2$ surface because its absolute value of local rotation is much lower than that using $\omega_{\perp e}$.
This suggests that resonant harmonic with plasma response mainly depends on absolute value of rotation near a rational surface rather than the existence of zero-crossing.
It is consistent with DIII-D experiments that absolute value of $\omega_{\perp e}$ is low during ELM suppression, and suppression lost is always accompanied with inward shift of $\omega_{\perp e}$ zero-crossing \cite{paz2019effect}, which could be interpreted by that when the absolute value of rotation is far away from zero at a rational surface, it has a strong shielding effect on RMP.

\subsection{Modeling of response in magnetic sensors}
\begin{figure}[htbp]
  \centering
  \includegraphics[width=0.48\textwidth]{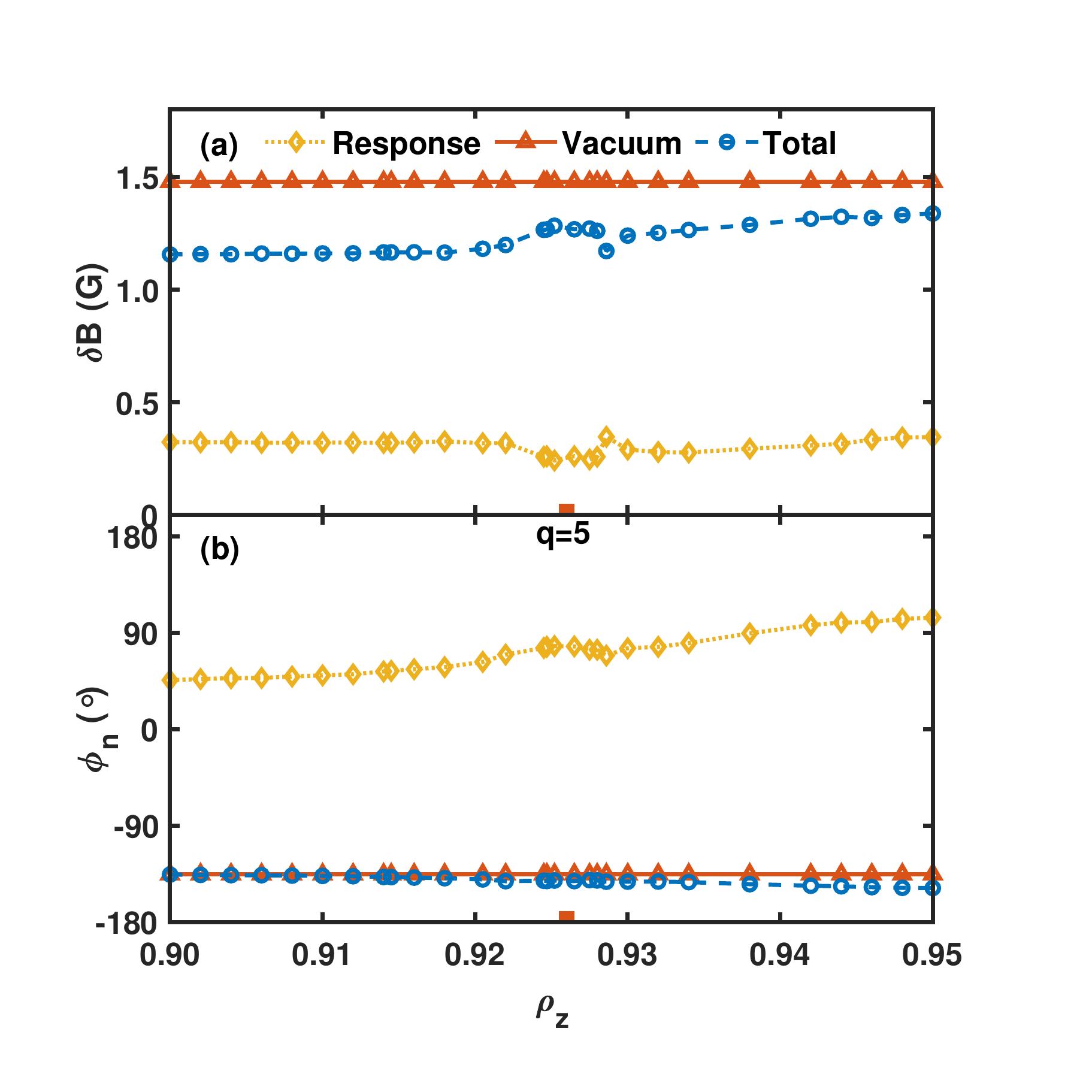}
  \vspace{-5mm}
  \caption{Dependences of amplitudes (a) and phase (b) of total $n=1$ magnetic perturbations (circles with dashed line) as well as vacuum (triangles with solid line) and pure response (diamonds with dotted line) components on rotation zero-crossing simulated by the MARS-F code.}
  \label{mfig:BPhi-HFS-U}
\end{figure}

There are several arrays of magnetic probes surrounding the vacuum chamber along the poloidal direction at high-field and low-field side in EAST \cite{ren2021penetration}.
Magnetic field measured by those probes at the high-field side is simulated, and dependence of its $n=1$ Fourier components on rotation zero-crossing is shown as figure \ref{mfig:BPhi-HFS-U} with total $n=1$ magnetic perturbations, vacuum and pure response are represented as circles with dashed line, triangles with solid line and diamonds with dotted line.
Both amplitude (a) and phase (b) of the response field show a small variation as rotation is nearly zero at $q=5$ rational surface, while no obvious changes in other region.
It shows that the increased tendency of $m/n=5/1$ resonant harmonic is not captured by plasma response field measured by magnetic probes, because there are also important contributions from other poloidal harmonics.
It might be a challenge to subtract this small variation in the measurement.
This suggests that the jump in the phase of response field during the transition from mitigation to suppression observed in previous EAST experiments \cite{sun2016nonlinear} indicates a nonlinear bifurcation in plasma response.

\section{Conclusions}
\label{sec:conclusion}
Plasma response to RMP at the pedestal top is of importance for ELM suppression in tokamak.
Rotation zero-crossing plays a key role in determining the threshold for field penetration of RMP.
Plasma response may be different for the case of low input torque plasma.
This work focuses on the dependence of plasma response to RMP on rotation zero-crossing based on linear MHD modeling using the MARS-F code and a reference EAST equilibrium.
It is shown that the plasma response is enhanced when zero-crossing occurs near rational surfaces.
The dependence of plasma response on the radial location of rotation zero-crossing is well fitted with a double Gaussian, indicating two effects in this enhancement.
One is induced by rotation screening effect shown as a wide base with a width around $10-20~\mathrm{krad/s}$, and the other is related to resistive singular layer effect characterized by a localized peak with a width around $3-4~\mathrm{krad/s}$.
The peak of the resonant harmonic in plasma response appears in the case where the rotation zero-crossing aligns with the rational surface, while the enhancement of magnetic islands' width at the peak is only around a factor of two.
Effects of plasma density, resistivity and rotation shear on the dependence are investigated for understanding the localized peak and the wide base profiles.
It is concluded that
1). The enhancement of resonant harmonic has no obvious relationship to the Alfv\'en resonance.
2). The width of the peak scales well with that of the resistive singular layer.
3). Rotation shear enhances the screening effect when the rotation crosses zero at the rational surface, and plasma response mainly depends on the absolute value of local rotation when the rotation zero-crossing is outside the singular layer.
These results are consistent with the radial displacement which is tearing type when zero-crossing is within the resistive singular layer, while it is kink type when zero-crossing is far from the layer.
Plasma responses at pedestal top for rotation profiles with or without zero-crossing are compared.
Similar width of the magnetic island induced by them suggests that the electron perpendicular rotation near pedestal top should not be far away from zero but may not be necessary to have zero-crossing for field penetration and hence accessing ELM suppression.
Simulated measurement of magnetic sensors for low rotation around the rational surface is also predicted for future experiments.

The results shown in this paper suggest that the rotation profile without zero-crossing may be acceptable for achieving ELM suppression if the rotation near the pedestal top is not far away from zero.
The modeling result shows the distance is of the order $10~\mathrm{krad/s}$.
If the distance exceeds this value, ELM suppression may be lost due to strong rotation shielding effect.
This is consistent with the correlation between location of $\omega_{\perp e}$ zero-crossing and ELM control effect in DIII-D experiments.
Besides, with lower plasma density, rotation shear and higher resistivity, the window of absolute value of rotation around rational surfaces would be broader for ELM suppression.
This work reveals detailed characteristics of plasma response to RMP in low input torque plasmas and is helpful for understanding the correlation between rotation zero-crossing and ELM suppression.
\section*{Acknowledgments}
\label{sec:acknowledgments}
This work is supported by the National Key R\&D Program of China under Grant No. 2017YFE0301100 and the National Natural Science Foundation of China under Grant No. 11875292 and No. 12005261.

\bibliography{aipsamp}

\providecommand{\noopsort}[1]{}\providecommand{\singleletter}[1]{#1}%
\begin{thebibliography}{46}%
\makeatletter
\providecommand \@ifxundefined [1]{%
 \@ifx{#1\undefined}
}%
\providecommand \@ifnum [1]{%
 \ifnum #1\expandafter \@firstoftwo
 \else \expandafter \@secondoftwo
 \fi
}%
\providecommand \@ifx [1]{%
 \ifx #1\expandafter \@firstoftwo
 \else \expandafter \@secondoftwo
 \fi
}%
\providecommand \natexlab [1]{#1}%
\providecommand \enquote  [1]{``#1''}%
\providecommand \bibnamefont  [1]{#1}%
\providecommand \bibfnamefont [1]{#1}%
\providecommand \citenamefont [1]{#1}%
\providecommand \href@noop [0]{\@secondoftwo}%
\providecommand \href [0]{\begingroup \@sanitize@url \@href}%
\providecommand \@href[1]{\@@startlink{#1}\@@href}%
\providecommand \@@href[1]{\endgroup#1\@@endlink}%
\providecommand \@sanitize@url [0]{\catcode `\\12\catcode `\$12\catcode
  `\&12\catcode `\#12\catcode `\^12\catcode `\_12\catcode `\%12\relax}%
\providecommand \@@startlink[1]{}%
\providecommand \@@endlink[0]{}%
\providecommand \url  [0]{\begingroup\@sanitize@url \@url }%
\providecommand \@url [1]{\endgroup\@href {#1}{\urlprefix }}%
\providecommand \urlprefix  [0]{URL }%
\providecommand \Eprint [0]{\href }%
\providecommand \doibase [0]{http://dx.doi.org/}%
\providecommand \selectlanguage [0]{\@gobble}%
\providecommand \bibinfo  [0]{\@secondoftwo}%
\providecommand \bibfield  [0]{\@secondoftwo}%
\providecommand \translation [1]{[#1]}%
\providecommand \BibitemOpen [0]{}%
\providecommand \bibitemStop [0]{}%
\providecommand \bibitemNoStop [0]{.\EOS\space}%
\providecommand \EOS [0]{\spacefactor3000\relax}%
\providecommand \BibitemShut  [1]{\csname bibitem#1\endcsname}%
\let\auto@bib@innerbib\@empty
\bibitem [{\citenamefont {Evans}(2015)}]{evans2015resonant}%
  \BibitemOpen
  \bibfield  {author} {\bibinfo {author} {\bibfnamefont {T.}~\bibnamefont
  {Evans}},\ }\bibfield  {title} {\enquote {\bibinfo {title} {Resonant magnetic
  perturbations of edge-plasmas in toroidal confinement devices},}\ }\href@noop
  {} {\bibfield  {journal} {\bibinfo  {journal} {Plasma Physics and Controlled
  Fusion}\ }\textbf {\bibinfo {volume} {57}},\ \bibinfo {pages} {123001}
  (\bibinfo {year} {2015})}\BibitemShut {NoStop}%
\bibitem [{\citenamefont {Loarte}\ \emph {et~al.}(2003)\citenamefont {Loarte},
  \citenamefont {Saibene}, \citenamefont {Sartori}, \citenamefont {Campbell},
  \citenamefont {Becoulet}, \citenamefont {Horton}, \citenamefont {Eich},
  \citenamefont {Herrmann}, \citenamefont {Matthews}, \citenamefont {Asakura}
  \emph {et~al.}}]{loarte2003characteristics}%
  \BibitemOpen
  \bibfield  {author} {\bibinfo {author} {\bibfnamefont {A.}~\bibnamefont
  {Loarte}}, \bibinfo {author} {\bibfnamefont {G.}~\bibnamefont {Saibene}},
  \bibinfo {author} {\bibfnamefont {R.}~\bibnamefont {Sartori}}, \bibinfo
  {author} {\bibfnamefont {D.}~\bibnamefont {Campbell}}, \bibinfo {author}
  {\bibfnamefont {M.}~\bibnamefont {Becoulet}}, \bibinfo {author}
  {\bibfnamefont {L.}~\bibnamefont {Horton}}, \bibinfo {author} {\bibfnamefont
  {T.}~\bibnamefont {Eich}}, \bibinfo {author} {\bibfnamefont {A.}~\bibnamefont
  {Herrmann}}, \bibinfo {author} {\bibfnamefont {G.}~\bibnamefont {Matthews}},
  \bibinfo {author} {\bibfnamefont {N.}~\bibnamefont {Asakura}},  \emph
  {et~al.},\ }\bibfield  {title} {\enquote {\bibinfo {title} {Characteristics
  of type {I} {ELM} energy and particle losses in existing devices and their
  extrapolation to {ITER}},}\ }\href@noop {} {\bibfield  {journal} {\bibinfo
  {journal} {Plasma Physics and Controlled Fusion}\ }\textbf {\bibinfo {volume}
  {45}},\ \bibinfo {pages} {1549} (\bibinfo {year} {2003})}\BibitemShut
  {NoStop}%
\bibitem [{\citenamefont {Evans}\ \emph {et~al.}(2005)\citenamefont {Evans},
  \citenamefont {Moyer}, \citenamefont {Watkins}, \citenamefont {Thomas},
  \citenamefont {Osborne}, \citenamefont {Boedo}, \citenamefont
  {Fenstermacher}, \citenamefont {Finken}, \citenamefont {Groebner},
  \citenamefont {Groth} \emph {et~al.}}]{evans2005suppression}%
  \BibitemOpen
  \bibfield  {author} {\bibinfo {author} {\bibfnamefont {T.}~\bibnamefont
  {Evans}}, \bibinfo {author} {\bibfnamefont {R.}~\bibnamefont {Moyer}},
  \bibinfo {author} {\bibfnamefont {J.}~\bibnamefont {Watkins}}, \bibinfo
  {author} {\bibfnamefont {P.}~\bibnamefont {Thomas}}, \bibinfo {author}
  {\bibfnamefont {T.}~\bibnamefont {Osborne}}, \bibinfo {author} {\bibfnamefont
  {J.}~\bibnamefont {Boedo}}, \bibinfo {author} {\bibfnamefont
  {M.}~\bibnamefont {Fenstermacher}}, \bibinfo {author} {\bibfnamefont
  {K.}~\bibnamefont {Finken}}, \bibinfo {author} {\bibfnamefont
  {R.}~\bibnamefont {Groebner}}, \bibinfo {author} {\bibfnamefont
  {M.}~\bibnamefont {Groth}},  \emph {et~al.},\ }\bibfield  {title} {\enquote
  {\bibinfo {title} {Suppression of large edge localized modes in high
  confinement {DIII-D} plasmas with a stochastic magnetic boundary},}\
  }\href@noop {} {\bibfield  {journal} {\bibinfo  {journal} {Journal of nuclear
  materials}\ }\textbf {\bibinfo {volume} {337}},\ \bibinfo {pages} {691--696}
  (\bibinfo {year} {2005})}\BibitemShut {NoStop}%
\bibitem [{\citenamefont {Jeon}\ \emph {et~al.}(2012)\citenamefont {Jeon},
  \citenamefont {Park}, \citenamefont {Yoon}, \citenamefont {Ko}, \citenamefont
  {Lee}, \citenamefont {Lee}, \citenamefont {Yun}, \citenamefont {Nam},
  \citenamefont {Kim}, \citenamefont {Kwak} \emph
  {et~al.}}]{jeon2012suppression}%
  \BibitemOpen
  \bibfield  {author} {\bibinfo {author} {\bibfnamefont {Y.}~\bibnamefont
  {Jeon}}, \bibinfo {author} {\bibfnamefont {J.-K.}\ \bibnamefont {Park}},
  \bibinfo {author} {\bibfnamefont {S.}~\bibnamefont {Yoon}}, \bibinfo {author}
  {\bibfnamefont {W.}~\bibnamefont {Ko}}, \bibinfo {author} {\bibfnamefont
  {S.}~\bibnamefont {Lee}}, \bibinfo {author} {\bibfnamefont {K.}~\bibnamefont
  {Lee}}, \bibinfo {author} {\bibfnamefont {G.}~\bibnamefont {Yun}}, \bibinfo
  {author} {\bibfnamefont {Y.}~\bibnamefont {Nam}}, \bibinfo {author}
  {\bibfnamefont {W.}~\bibnamefont {Kim}}, \bibinfo {author} {\bibfnamefont
  {J.-G.}\ \bibnamefont {Kwak}},  \emph {et~al.},\ }\bibfield  {title}
  {\enquote {\bibinfo {title} {Suppression of edge localized modes in
  high-confinement {KSTAR} plasmas by nonaxisymmetric magnetic
  perturbations},}\ }\href@noop {} {\bibfield  {journal} {\bibinfo  {journal}
  {Physical review letters}\ }\textbf {\bibinfo {volume} {109}},\ \bibinfo
  {pages} {035004} (\bibinfo {year} {2012})}\BibitemShut {NoStop}%
\bibitem [{\citenamefont {Sun}\ \emph {et~al.}(2016{\natexlab{a}})\citenamefont
  {Sun}, \citenamefont {Liang}, \citenamefont {Liu}, \citenamefont {Gu},
  \citenamefont {Yang}, \citenamefont {Guo}, \citenamefont {Shi}, \citenamefont
  {Jia}, \citenamefont {Wang}, \citenamefont {Lyu} \emph
  {et~al.}}]{sun2016nonlinear}%
  \BibitemOpen
  \bibfield  {author} {\bibinfo {author} {\bibfnamefont {Y.}~\bibnamefont
  {Sun}}, \bibinfo {author} {\bibfnamefont {Y.}~\bibnamefont {Liang}}, \bibinfo
  {author} {\bibfnamefont {Y.}~\bibnamefont {Liu}}, \bibinfo {author}
  {\bibfnamefont {S.}~\bibnamefont {Gu}}, \bibinfo {author} {\bibfnamefont
  {X.}~\bibnamefont {Yang}}, \bibinfo {author} {\bibfnamefont {W.}~\bibnamefont
  {Guo}}, \bibinfo {author} {\bibfnamefont {T.}~\bibnamefont {Shi}}, \bibinfo
  {author} {\bibfnamefont {M.}~\bibnamefont {Jia}}, \bibinfo {author}
  {\bibfnamefont {L.}~\bibnamefont {Wang}}, \bibinfo {author} {\bibfnamefont
  {B.}~\bibnamefont {Lyu}},  \emph {et~al.},\ }\bibfield  {title} {\enquote
  {\bibinfo {title} {Nonlinear transition from mitigation to suppression of the
  edge localized mode with resonant magnetic perturbations in the {EAST}
  tokamak},}\ }\href@noop {} {\bibfield  {journal} {\bibinfo  {journal}
  {Physical review letters}\ }\textbf {\bibinfo {volume} {117}},\ \bibinfo
  {pages} {115001} (\bibinfo {year} {2016}{\natexlab{a}})}\BibitemShut
  {NoStop}%
\bibitem [{\citenamefont {Suttrop}\ \emph {et~al.}(2018)\citenamefont
  {Suttrop}, \citenamefont {Kirk}, \citenamefont {Bobkov}, \citenamefont
  {Cavedon}, \citenamefont {Dunne}, \citenamefont {McDermott}, \citenamefont
  {Meyer}, \citenamefont {Nazikian}, \citenamefont {Paz-Soldan}, \citenamefont
  {Ryan} \emph {et~al.}}]{suttrop2018experimental}%
  \BibitemOpen
  \bibfield  {author} {\bibinfo {author} {\bibfnamefont {W.}~\bibnamefont
  {Suttrop}}, \bibinfo {author} {\bibfnamefont {A.}~\bibnamefont {Kirk}},
  \bibinfo {author} {\bibfnamefont {V.}~\bibnamefont {Bobkov}}, \bibinfo
  {author} {\bibfnamefont {M.}~\bibnamefont {Cavedon}}, \bibinfo {author}
  {\bibfnamefont {M.}~\bibnamefont {Dunne}}, \bibinfo {author} {\bibfnamefont
  {R.}~\bibnamefont {McDermott}}, \bibinfo {author} {\bibfnamefont
  {H.}~\bibnamefont {Meyer}}, \bibinfo {author} {\bibfnamefont
  {R.}~\bibnamefont {Nazikian}}, \bibinfo {author} {\bibfnamefont
  {C.}~\bibnamefont {Paz-Soldan}}, \bibinfo {author} {\bibfnamefont {D.~A.}\
  \bibnamefont {Ryan}},  \emph {et~al.},\ }\bibfield  {title} {\enquote
  {\bibinfo {title} {Experimental conditions to suppress edge localised modes
  by magnetic perturbations in the {ASDEX} {Upgrade} tokamak},}\ }\href@noop {}
  {\bibfield  {journal} {\bibinfo  {journal} {Nuclear Fusion}\ }\textbf
  {\bibinfo {volume} {58}},\ \bibinfo {pages} {096031} (\bibinfo {year}
  {2018})}\BibitemShut {NoStop}%
\bibitem [{\citenamefont {Liang}\ \emph {et~al.}(2007)\citenamefont {Liang},
  \citenamefont {Koslowski}, \citenamefont {Thomas}, \citenamefont {Nardon},
  \citenamefont {Alper}, \citenamefont {Andrew}, \citenamefont {Andrew},
  \citenamefont {Arnoux}, \citenamefont {Baranov}, \citenamefont {B{\'e}coulet}
  \emph {et~al.}}]{liang2007active}%
  \BibitemOpen
  \bibfield  {author} {\bibinfo {author} {\bibfnamefont {Y.}~\bibnamefont
  {Liang}}, \bibinfo {author} {\bibfnamefont {H.}~\bibnamefont {Koslowski}},
  \bibinfo {author} {\bibfnamefont {P.}~\bibnamefont {Thomas}}, \bibinfo
  {author} {\bibfnamefont {E.}~\bibnamefont {Nardon}}, \bibinfo {author}
  {\bibfnamefont {B.}~\bibnamefont {Alper}}, \bibinfo {author} {\bibfnamefont
  {P.}~\bibnamefont {Andrew}}, \bibinfo {author} {\bibfnamefont
  {Y.}~\bibnamefont {Andrew}}, \bibinfo {author} {\bibfnamefont
  {G.}~\bibnamefont {Arnoux}}, \bibinfo {author} {\bibfnamefont
  {Y.}~\bibnamefont {Baranov}}, \bibinfo {author} {\bibfnamefont
  {M.}~\bibnamefont {B{\'e}coulet}},  \emph {et~al.},\ }\bibfield  {title}
  {\enquote {\bibinfo {title} {Active control of type-i edge-localized modes
  with n= 1 perturbation fields in the {JET} tokamak},}\ }\href@noop {}
  {\bibfield  {journal} {\bibinfo  {journal} {Physical review letters}\
  }\textbf {\bibinfo {volume} {98}},\ \bibinfo {pages} {265004} (\bibinfo
  {year} {2007})}\BibitemShut {NoStop}%
\bibitem [{\citenamefont {Kirk}\ \emph {et~al.}(2013)\citenamefont {Kirk},
  \citenamefont {Chapman}, \citenamefont {Liu}, \citenamefont {Cahyna},
  \citenamefont {Denner}, \citenamefont {Fishpool}, \citenamefont {Ham},
  \citenamefont {Harrison}, \citenamefont {Liang}, \citenamefont {Nardon} \emph
  {et~al.}}]{kirk2013understanding}%
  \BibitemOpen
  \bibfield  {author} {\bibinfo {author} {\bibfnamefont {A.}~\bibnamefont
  {Kirk}}, \bibinfo {author} {\bibfnamefont {I.}~\bibnamefont {Chapman}},
  \bibinfo {author} {\bibfnamefont {Y.}~\bibnamefont {Liu}}, \bibinfo {author}
  {\bibfnamefont {P.}~\bibnamefont {Cahyna}}, \bibinfo {author} {\bibfnamefont
  {P.}~\bibnamefont {Denner}}, \bibinfo {author} {\bibfnamefont
  {G.}~\bibnamefont {Fishpool}}, \bibinfo {author} {\bibfnamefont
  {C.}~\bibnamefont {Ham}}, \bibinfo {author} {\bibfnamefont {J.}~\bibnamefont
  {Harrison}}, \bibinfo {author} {\bibfnamefont {Y.}~\bibnamefont {Liang}},
  \bibinfo {author} {\bibfnamefont {E.}~\bibnamefont {Nardon}},  \emph
  {et~al.},\ }\bibfield  {title} {\enquote {\bibinfo {title} {Understanding
  edge-localized mode mitigation by resonant magnetic perturbations on
  {MAST}},}\ }\href@noop {} {\bibfield  {journal} {\bibinfo  {journal} {Nuclear
  Fusion}\ }\textbf {\bibinfo {volume} {53}},\ \bibinfo {pages} {043007}
  (\bibinfo {year} {2013})}\BibitemShut {NoStop}%
\bibitem [{\citenamefont {Liu}, \citenamefont {Kirk},\ and\ \citenamefont
  {Nardon}(2010)}]{liu2010full}%
  \BibitemOpen
  \bibfield  {author} {\bibinfo {author} {\bibfnamefont {Y.}~\bibnamefont
  {Liu}}, \bibinfo {author} {\bibfnamefont {A.}~\bibnamefont {Kirk}}, \ and\
  \bibinfo {author} {\bibfnamefont {E.}~\bibnamefont {Nardon}},\ }\bibfield
  {title} {\enquote {\bibinfo {title} {Full toroidal plasma response to
  externally applied nonaxisymmetric magnetic fields},}\ }\href@noop {}
  {\bibfield  {journal} {\bibinfo  {journal} {Physics of Plasmas}\ }\textbf
  {\bibinfo {volume} {17}},\ \bibinfo {pages} {122502} (\bibinfo {year}
  {2010})}\BibitemShut {NoStop}%
\bibitem [{\citenamefont {Paz-Soldan}\ \emph {et~al.}(2015)\citenamefont
  {Paz-Soldan}, \citenamefont {Nazikian}, \citenamefont {Haskey}, \citenamefont
  {Logan}, \citenamefont {Strait}, \citenamefont {Ferraro}, \citenamefont
  {Hanson}, \citenamefont {King}, \citenamefont {Lanctot}, \citenamefont
  {Moyer} \emph {et~al.}}]{paz2015observation}%
  \BibitemOpen
  \bibfield  {author} {\bibinfo {author} {\bibfnamefont {C.}~\bibnamefont
  {Paz-Soldan}}, \bibinfo {author} {\bibfnamefont {R.}~\bibnamefont
  {Nazikian}}, \bibinfo {author} {\bibfnamefont {S.}~\bibnamefont {Haskey}},
  \bibinfo {author} {\bibfnamefont {N.}~\bibnamefont {Logan}}, \bibinfo
  {author} {\bibfnamefont {E.}~\bibnamefont {Strait}}, \bibinfo {author}
  {\bibfnamefont {N.}~\bibnamefont {Ferraro}}, \bibinfo {author} {\bibfnamefont
  {J.}~\bibnamefont {Hanson}}, \bibinfo {author} {\bibfnamefont
  {J.}~\bibnamefont {King}}, \bibinfo {author} {\bibfnamefont {M.}~\bibnamefont
  {Lanctot}}, \bibinfo {author} {\bibfnamefont {R.}~\bibnamefont {Moyer}},
  \emph {et~al.},\ }\bibfield  {title} {\enquote {\bibinfo {title} {Observation
  of a multimode plasma response and its relationship to density pumpout and
  edge-localized mode suppression},}\ }\href@noop {} {\bibfield  {journal}
  {\bibinfo  {journal} {Physical review letters}\ }\textbf {\bibinfo {volume}
  {114}},\ \bibinfo {pages} {105001} (\bibinfo {year} {2015})}\BibitemShut
  {NoStop}%
\bibitem [{\citenamefont {Liu}\ \emph {et~al.}(2016)\citenamefont {Liu},
  \citenamefont {Ham}, \citenamefont {Kirk}, \citenamefont {Li}, \citenamefont
  {Loarte}, \citenamefont {Ryan}, \citenamefont {Sun}, \citenamefont {Suttrop},
  \citenamefont {Yang},\ and\ \citenamefont {Zhou}}]{liu2016elm}%
  \BibitemOpen
  \bibfield  {author} {\bibinfo {author} {\bibfnamefont {Y.}~\bibnamefont
  {Liu}}, \bibinfo {author} {\bibfnamefont {C.}~\bibnamefont {Ham}}, \bibinfo
  {author} {\bibfnamefont {A.}~\bibnamefont {Kirk}}, \bibinfo {author}
  {\bibfnamefont {L.}~\bibnamefont {Li}}, \bibinfo {author} {\bibfnamefont
  {A.}~\bibnamefont {Loarte}}, \bibinfo {author} {\bibfnamefont
  {D.}~\bibnamefont {Ryan}}, \bibinfo {author} {\bibfnamefont {Y.}~\bibnamefont
  {Sun}}, \bibinfo {author} {\bibfnamefont {W.}~\bibnamefont {Suttrop}},
  \bibinfo {author} {\bibfnamefont {X.}~\bibnamefont {Yang}}, \ and\ \bibinfo
  {author} {\bibfnamefont {L.}~\bibnamefont {Zhou}},\ }\bibfield  {title}
  {\enquote {\bibinfo {title} {Elm control with {RMP}: Plasma response models
  and the role of edge peeling response},}\ }\href@noop {} {\bibfield
  {journal} {\bibinfo  {journal} {Plasma Physics and Controlled Fusion}\
  }\textbf {\bibinfo {volume} {58}},\ \bibinfo {pages} {114005} (\bibinfo
  {year} {2016})}\BibitemShut {NoStop}%
\bibitem [{\citenamefont {Yang}\ \emph {et~al.}(2016)\citenamefont {Yang},
  \citenamefont {Sun}, \citenamefont {Liu}, \citenamefont {Gu}, \citenamefont
  {Liu}, \citenamefont {Wang}, \citenamefont {Zhou},\ and\ \citenamefont
  {Guo}}]{yang2016modelling}%
  \BibitemOpen
  \bibfield  {author} {\bibinfo {author} {\bibfnamefont {X.}~\bibnamefont
  {Yang}}, \bibinfo {author} {\bibfnamefont {Y.}~\bibnamefont {Sun}}, \bibinfo
  {author} {\bibfnamefont {Y.}~\bibnamefont {Liu}}, \bibinfo {author}
  {\bibfnamefont {S.}~\bibnamefont {Gu}}, \bibinfo {author} {\bibfnamefont
  {Y.}~\bibnamefont {Liu}}, \bibinfo {author} {\bibfnamefont {H.}~\bibnamefont
  {Wang}}, \bibinfo {author} {\bibfnamefont {L.}~\bibnamefont {Zhou}}, \ and\
  \bibinfo {author} {\bibfnamefont {W.}~\bibnamefont {Guo}},\ }\bibfield
  {title} {\enquote {\bibinfo {title} {Modelling of plasma response to {3D}
  external magnetic field perturbations in {EAST}},}\ }\href@noop {} {\bibfield
   {journal} {\bibinfo  {journal} {Plasma Physics and Controlled Fusion}\
  }\textbf {\bibinfo {volume} {58}},\ \bibinfo {pages} {114006} (\bibinfo
  {year} {2016})}\BibitemShut {NoStop}%
\bibitem [{\citenamefont {Sun}\ \emph {et~al.}(2016{\natexlab{b}})\citenamefont
  {Sun}, \citenamefont {Jia}, \citenamefont {Zang}, \citenamefont {Wang},
  \citenamefont {Liang}, \citenamefont {Liu}, \citenamefont {Yang},
  \citenamefont {Guo}, \citenamefont {Gu}, \citenamefont {Li} \emph
  {et~al.}}]{sun2016edge}%
  \BibitemOpen
  \bibfield  {author} {\bibinfo {author} {\bibfnamefont {Y.}~\bibnamefont
  {Sun}}, \bibinfo {author} {\bibfnamefont {M.}~\bibnamefont {Jia}}, \bibinfo
  {author} {\bibfnamefont {Q.}~\bibnamefont {Zang}}, \bibinfo {author}
  {\bibfnamefont {L.}~\bibnamefont {Wang}}, \bibinfo {author} {\bibfnamefont
  {Y.}~\bibnamefont {Liang}}, \bibinfo {author} {\bibfnamefont
  {Y.}~\bibnamefont {Liu}}, \bibinfo {author} {\bibfnamefont {X.}~\bibnamefont
  {Yang}}, \bibinfo {author} {\bibfnamefont {W.}~\bibnamefont {Guo}}, \bibinfo
  {author} {\bibfnamefont {S.}~\bibnamefont {Gu}}, \bibinfo {author}
  {\bibfnamefont {Y.}~\bibnamefont {Li}},  \emph {et~al.},\ }\bibfield  {title}
  {\enquote {\bibinfo {title} {Edge localized mode control using n= 1 resonant
  magnetic perturbation in the {EAST} tokamak},}\ }\href@noop {} {\bibfield
  {journal} {\bibinfo  {journal} {Nuclear Fusion}\ }\textbf {\bibinfo {volume}
  {57}},\ \bibinfo {pages} {036007} (\bibinfo {year}
  {2016}{\natexlab{b}})}\BibitemShut {NoStop}%
\bibitem [{\citenamefont {Park}\ \emph {et~al.}(2018)\citenamefont {Park},
  \citenamefont {Jeon}, \citenamefont {In}, \citenamefont {Ahn}, \citenamefont
  {Nazikian}, \citenamefont {Park}, \citenamefont {Kim}, \citenamefont {Lee},
  \citenamefont {Ko}, \citenamefont {Kim} \emph {et~al.}}]{park20183d}%
  \BibitemOpen
  \bibfield  {author} {\bibinfo {author} {\bibfnamefont {J.-K.}\ \bibnamefont
  {Park}}, \bibinfo {author} {\bibfnamefont {Y.}~\bibnamefont {Jeon}}, \bibinfo
  {author} {\bibfnamefont {Y.}~\bibnamefont {In}}, \bibinfo {author}
  {\bibfnamefont {J.-W.}\ \bibnamefont {Ahn}}, \bibinfo {author} {\bibfnamefont
  {R.}~\bibnamefont {Nazikian}}, \bibinfo {author} {\bibfnamefont
  {G.}~\bibnamefont {Park}}, \bibinfo {author} {\bibfnamefont {J.}~\bibnamefont
  {Kim}}, \bibinfo {author} {\bibfnamefont {H.}~\bibnamefont {Lee}}, \bibinfo
  {author} {\bibfnamefont {W.}~\bibnamefont {Ko}}, \bibinfo {author}
  {\bibfnamefont {H.-S.}\ \bibnamefont {Kim}},  \emph {et~al.},\ }\bibfield
  {title} {\enquote {\bibinfo {title} {3d field phase-space control in tokamak
  plasmas},}\ }\href@noop {} {\bibfield  {journal} {\bibinfo  {journal} {Nature
  Physics}\ }\textbf {\bibinfo {volume} {14}},\ \bibinfo {pages} {1223--1228}
  (\bibinfo {year} {2018})}\BibitemShut {NoStop}%
\bibitem [{\citenamefont {Gu}\ \emph {et~al.}(2019)\citenamefont {Gu},
  \citenamefont {Wan}, \citenamefont {Sun}, \citenamefont {Chu}, \citenamefont
  {Liu}, \citenamefont {Shi}, \citenamefont {Wang}, \citenamefont {Jia},\ and\
  \citenamefont {He}}]{gu2019new}%
  \BibitemOpen
  \bibfield  {author} {\bibinfo {author} {\bibfnamefont {S.}~\bibnamefont
  {Gu}}, \bibinfo {author} {\bibfnamefont {B.}~\bibnamefont {Wan}}, \bibinfo
  {author} {\bibfnamefont {Y.}~\bibnamefont {Sun}}, \bibinfo {author}
  {\bibfnamefont {N.}~\bibnamefont {Chu}}, \bibinfo {author} {\bibfnamefont
  {Y.}~\bibnamefont {Liu}}, \bibinfo {author} {\bibfnamefont {T.}~\bibnamefont
  {Shi}}, \bibinfo {author} {\bibfnamefont {H.}~\bibnamefont {Wang}}, \bibinfo
  {author} {\bibfnamefont {M.}~\bibnamefont {Jia}}, \ and\ \bibinfo {author}
  {\bibfnamefont {K.}~\bibnamefont {He}},\ }\bibfield  {title} {\enquote
  {\bibinfo {title} {A new criterion for controlling edge localized modes based
  on a multi-mode plasma response},}\ }\href@noop {} {\bibfield  {journal}
  {\bibinfo  {journal} {Nuclear Fusion}\ }\textbf {\bibinfo {volume} {59}},\
  \bibinfo {pages} {126042} (\bibinfo {year} {2019})}\BibitemShut {NoStop}%
\bibitem [{\citenamefont {Snyder}\ \emph {et~al.}(2012)\citenamefont {Snyder},
  \citenamefont {Osborne}, \citenamefont {Burrell}, \citenamefont {Groebner},
  \citenamefont {Leonard}, \citenamefont {Nazikian}, \citenamefont {Orlov},
  \citenamefont {Schmitz}, \citenamefont {Wade},\ and\ \citenamefont
  {Wilson}}]{snyder2012eped}%
  \BibitemOpen
  \bibfield  {author} {\bibinfo {author} {\bibfnamefont {P.}~\bibnamefont
  {Snyder}}, \bibinfo {author} {\bibfnamefont {T.}~\bibnamefont {Osborne}},
  \bibinfo {author} {\bibfnamefont {K.}~\bibnamefont {Burrell}}, \bibinfo
  {author} {\bibfnamefont {R.}~\bibnamefont {Groebner}}, \bibinfo {author}
  {\bibfnamefont {A.}~\bibnamefont {Leonard}}, \bibinfo {author} {\bibfnamefont
  {R.}~\bibnamefont {Nazikian}}, \bibinfo {author} {\bibfnamefont
  {D.}~\bibnamefont {Orlov}}, \bibinfo {author} {\bibfnamefont
  {O.}~\bibnamefont {Schmitz}}, \bibinfo {author} {\bibfnamefont
  {M.}~\bibnamefont {Wade}}, \ and\ \bibinfo {author} {\bibfnamefont
  {H.}~\bibnamefont {Wilson}},\ }\bibfield  {title} {\enquote {\bibinfo {title}
  {The {EPED} pedestal model and edge localized mode-suppressed regimes:
  Studies of quiescent {H-mode} and development of a model for edge localized
  mode suppression via resonant magnetic perturbations},}\ }\href@noop {}
  {\bibfield  {journal} {\bibinfo  {journal} {Physics of plasmas}\ }\textbf
  {\bibinfo {volume} {19}},\ \bibinfo {pages} {056115} (\bibinfo {year}
  {2012})}\BibitemShut {NoStop}%
\bibitem [{\citenamefont {Wade}\ \emph {et~al.}(2015)\citenamefont {Wade},
  \citenamefont {Nazikian}, \citenamefont {DeGrassie}, \citenamefont {Evans},
  \citenamefont {Ferraro}, \citenamefont {Moyer}, \citenamefont {Orlov},
  \citenamefont {Buttery}, \citenamefont {Fenstermacher}, \citenamefont
  {Garofalo} \emph {et~al.}}]{wade2015advances}%
  \BibitemOpen
  \bibfield  {author} {\bibinfo {author} {\bibfnamefont {M.}~\bibnamefont
  {Wade}}, \bibinfo {author} {\bibfnamefont {R.}~\bibnamefont {Nazikian}},
  \bibinfo {author} {\bibfnamefont {J.}~\bibnamefont {DeGrassie}}, \bibinfo
  {author} {\bibfnamefont {T.}~\bibnamefont {Evans}}, \bibinfo {author}
  {\bibfnamefont {N.}~\bibnamefont {Ferraro}}, \bibinfo {author} {\bibfnamefont
  {R.}~\bibnamefont {Moyer}}, \bibinfo {author} {\bibfnamefont
  {D.}~\bibnamefont {Orlov}}, \bibinfo {author} {\bibfnamefont
  {R.}~\bibnamefont {Buttery}}, \bibinfo {author} {\bibfnamefont
  {M.}~\bibnamefont {Fenstermacher}}, \bibinfo {author} {\bibfnamefont
  {A.}~\bibnamefont {Garofalo}},  \emph {et~al.},\ }\bibfield  {title}
  {\enquote {\bibinfo {title} {Advances in the physics understanding of {ELM}
  suppression using resonant magnetic perturbations in {DIII-D}},}\ }\href@noop
  {} {\bibfield  {journal} {\bibinfo  {journal} {Nuclear Fusion}\ }\textbf
  {\bibinfo {volume} {55}},\ \bibinfo {pages} {023002} (\bibinfo {year}
  {2015})}\BibitemShut {NoStop}%
\bibitem [{\citenamefont {Evans}\ \emph {et~al.}(2004)\citenamefont {Evans},
  \citenamefont {Moyer}, \citenamefont {Thomas}, \citenamefont {Watkins},
  \citenamefont {Osborne}, \citenamefont {Boedo}, \citenamefont {Doyle},
  \citenamefont {Fenstermacher}, \citenamefont {Finken}, \citenamefont
  {Groebner}, \citenamefont {Groth}, \citenamefont {Harris}, \citenamefont
  {La~Haye}, \citenamefont {Lasnier}, \citenamefont {Masuzaki}, \citenamefont
  {Ohyabu}, \citenamefont {Pretty}, \citenamefont {Rhodes}, \citenamefont
  {Reimerdes}, \citenamefont {Rudakov}, \citenamefont {Schaffer}, \citenamefont
  {Wang},\ and\ \citenamefont {Zeng}}]{evans2004suppression}%
  \BibitemOpen
  \bibfield  {author} {\bibinfo {author} {\bibfnamefont {T.~E.}\ \bibnamefont
  {Evans}}, \bibinfo {author} {\bibfnamefont {R.~A.}\ \bibnamefont {Moyer}},
  \bibinfo {author} {\bibfnamefont {P.~R.}\ \bibnamefont {Thomas}}, \bibinfo
  {author} {\bibfnamefont {J.~G.}\ \bibnamefont {Watkins}}, \bibinfo {author}
  {\bibfnamefont {T.~H.}\ \bibnamefont {Osborne}}, \bibinfo {author}
  {\bibfnamefont {J.~A.}\ \bibnamefont {Boedo}}, \bibinfo {author}
  {\bibfnamefont {E.~J.}\ \bibnamefont {Doyle}}, \bibinfo {author}
  {\bibfnamefont {M.~E.}\ \bibnamefont {Fenstermacher}}, \bibinfo {author}
  {\bibfnamefont {K.~H.}\ \bibnamefont {Finken}}, \bibinfo {author}
  {\bibfnamefont {R.~J.}\ \bibnamefont {Groebner}}, \bibinfo {author}
  {\bibfnamefont {M.}~\bibnamefont {Groth}}, \bibinfo {author} {\bibfnamefont
  {J.~H.}\ \bibnamefont {Harris}}, \bibinfo {author} {\bibfnamefont {R.~J.}\
  \bibnamefont {La~Haye}}, \bibinfo {author} {\bibfnamefont {C.~J.}\
  \bibnamefont {Lasnier}}, \bibinfo {author} {\bibfnamefont {S.}~\bibnamefont
  {Masuzaki}}, \bibinfo {author} {\bibfnamefont {N.}~\bibnamefont {Ohyabu}},
  \bibinfo {author} {\bibfnamefont {D.~G.}\ \bibnamefont {Pretty}}, \bibinfo
  {author} {\bibfnamefont {T.~L.}\ \bibnamefont {Rhodes}}, \bibinfo {author}
  {\bibfnamefont {H.}~\bibnamefont {Reimerdes}}, \bibinfo {author}
  {\bibfnamefont {D.~L.}\ \bibnamefont {Rudakov}}, \bibinfo {author}
  {\bibfnamefont {M.~J.}\ \bibnamefont {Schaffer}}, \bibinfo {author}
  {\bibfnamefont {G.}~\bibnamefont {Wang}}, \ and\ \bibinfo {author}
  {\bibfnamefont {L.}~\bibnamefont {Zeng}},\ }\bibfield  {title} {\enquote
  {\bibinfo {title} {Suppression of large edge-localized modes in
  high-confinement {DIII-D} plasmas with a stochastic magnetic boundary},}\
  }\href {\doibase 10.1103/PhysRevLett.92.235003} {\bibfield  {journal}
  {\bibinfo  {journal} {Phys. Rev. Lett.}\ }\textbf {\bibinfo {volume} {92}},\
  \bibinfo {pages} {235003} (\bibinfo {year} {2004})}\BibitemShut {NoStop}%
\bibitem [{\citenamefont {Fenstermacher}\ \emph {et~al.}(2008)\citenamefont
  {Fenstermacher}, \citenamefont {Evans}, \citenamefont {Osborne},
  \citenamefont {Schaffer}, \citenamefont {Aldan}, \citenamefont {Degrassie},
  \citenamefont {Gohil}, \citenamefont {Joseph}, \citenamefont {Moyer},
  \citenamefont {Snyder} \emph {et~al.}}]{fenstermacher2008effect}%
  \BibitemOpen
  \bibfield  {author} {\bibinfo {author} {\bibfnamefont {M.}~\bibnamefont
  {Fenstermacher}}, \bibinfo {author} {\bibfnamefont {T.}~\bibnamefont
  {Evans}}, \bibinfo {author} {\bibfnamefont {T.}~\bibnamefont {Osborne}},
  \bibinfo {author} {\bibfnamefont {M.}~\bibnamefont {Schaffer}}, \bibinfo
  {author} {\bibfnamefont {M.}~\bibnamefont {Aldan}}, \bibinfo {author}
  {\bibfnamefont {J.}~\bibnamefont {Degrassie}}, \bibinfo {author}
  {\bibfnamefont {P.}~\bibnamefont {Gohil}}, \bibinfo {author} {\bibfnamefont
  {I.}~\bibnamefont {Joseph}}, \bibinfo {author} {\bibfnamefont
  {R.}~\bibnamefont {Moyer}}, \bibinfo {author} {\bibfnamefont
  {P.}~\bibnamefont {Snyder}},  \emph {et~al.},\ }\bibfield  {title} {\enquote
  {\bibinfo {title} {Effect of island overlap on edge localized mode
  suppression by resonant magnetic perturbations in {DIII-D}},}\ }\href@noop {}
  {\bibfield  {journal} {\bibinfo  {journal} {Physics of Plasmas}\ }\textbf
  {\bibinfo {volume} {15}},\ \bibinfo {pages} {056122} (\bibinfo {year}
  {2008})}\BibitemShut {NoStop}%
\bibitem [{\citenamefont {Nazikian}\ \emph {et~al.}(2015)\citenamefont
  {Nazikian}, \citenamefont {Paz-Soldan}, \citenamefont {Callen}, \citenamefont
  {DeGrassie}, \citenamefont {Eldon}, \citenamefont {Evans}, \citenamefont
  {Ferraro}, \citenamefont {Grierson}, \citenamefont {Groebner}, \citenamefont
  {Haskey} \emph {et~al.}}]{nazikian2015pedestal}%
  \BibitemOpen
  \bibfield  {author} {\bibinfo {author} {\bibfnamefont {R.}~\bibnamefont
  {Nazikian}}, \bibinfo {author} {\bibfnamefont {C.}~\bibnamefont
  {Paz-Soldan}}, \bibinfo {author} {\bibfnamefont {J.}~\bibnamefont {Callen}},
  \bibinfo {author} {\bibfnamefont {J.}~\bibnamefont {DeGrassie}}, \bibinfo
  {author} {\bibfnamefont {D.}~\bibnamefont {Eldon}}, \bibinfo {author}
  {\bibfnamefont {T.}~\bibnamefont {Evans}}, \bibinfo {author} {\bibfnamefont
  {N.}~\bibnamefont {Ferraro}}, \bibinfo {author} {\bibfnamefont
  {B.}~\bibnamefont {Grierson}}, \bibinfo {author} {\bibfnamefont
  {R.}~\bibnamefont {Groebner}}, \bibinfo {author} {\bibfnamefont
  {S.}~\bibnamefont {Haskey}},  \emph {et~al.},\ }\bibfield  {title} {\enquote
  {\bibinfo {title} {Pedestal bifurcation and resonant field penetration at the
  threshold of edge-localized mode suppression in the {DIII-D} tokamak},}\
  }\href@noop {} {\bibfield  {journal} {\bibinfo  {journal} {Physical review
  letters}\ }\textbf {\bibinfo {volume} {114}},\ \bibinfo {pages} {105002}
  (\bibinfo {year} {2015})}\BibitemShut {NoStop}%
\bibitem [{\citenamefont {Hu}\ \emph {et~al.}(2020)\citenamefont {Hu},
  \citenamefont {Nazikian}, \citenamefont {Grierson}, \citenamefont {Logan},
  \citenamefont {Orlov}, \citenamefont {Paz-Soldan},\ and\ \citenamefont
  {Yu}}]{hu2020wide}%
  \BibitemOpen
  \bibfield  {author} {\bibinfo {author} {\bibfnamefont {Q.}~\bibnamefont
  {Hu}}, \bibinfo {author} {\bibfnamefont {R.}~\bibnamefont {Nazikian}},
  \bibinfo {author} {\bibfnamefont {B.}~\bibnamefont {Grierson}}, \bibinfo
  {author} {\bibfnamefont {N.}~\bibnamefont {Logan}}, \bibinfo {author}
  {\bibfnamefont {D.}~\bibnamefont {Orlov}}, \bibinfo {author} {\bibfnamefont
  {C.}~\bibnamefont {Paz-Soldan}}, \ and\ \bibinfo {author} {\bibfnamefont
  {Q.}~\bibnamefont {Yu}},\ }\bibfield  {title} {\enquote {\bibinfo {title}
  {Wide operational windows of edge-localized mode suppression by resonant
  magnetic perturbations in the {DIII-D} tokamak},}\ }\href@noop {} {\bibfield
  {journal} {\bibinfo  {journal} {Physical Review Letters}\ }\textbf {\bibinfo
  {volume} {125}},\ \bibinfo {pages} {045001} (\bibinfo {year}
  {2020})}\BibitemShut {NoStop}%
\bibitem [{\citenamefont {Fitzpatrick}(1998)}]{fitzpatrick1998bifurcated}%
  \BibitemOpen
  \bibfield  {author} {\bibinfo {author} {\bibfnamefont {R.}~\bibnamefont
  {Fitzpatrick}},\ }\bibfield  {title} {\enquote {\bibinfo {title} {Bifurcated
  states of a rotating tokamak plasma in the presence of a static
  error-field},}\ }\href@noop {} {\bibfield  {journal} {\bibinfo  {journal}
  {Physics of Plasmas}\ }\textbf {\bibinfo {volume} {5}},\ \bibinfo {pages}
  {3325--3341} (\bibinfo {year} {1998})}\BibitemShut {NoStop}%
\bibitem [{\citenamefont {Waelbroeck}\ \emph {et~al.}(2012)\citenamefont
  {Waelbroeck}, \citenamefont {Joseph}, \citenamefont {Nardon}, \citenamefont
  {B{\'e}coulet},\ and\ \citenamefont {Fitzpatrick}}]{waelbroeck2012role}%
  \BibitemOpen
  \bibfield  {author} {\bibinfo {author} {\bibfnamefont {F.}~\bibnamefont
  {Waelbroeck}}, \bibinfo {author} {\bibfnamefont {I.}~\bibnamefont {Joseph}},
  \bibinfo {author} {\bibfnamefont {E.}~\bibnamefont {Nardon}}, \bibinfo
  {author} {\bibfnamefont {M.}~\bibnamefont {B{\'e}coulet}}, \ and\ \bibinfo
  {author} {\bibfnamefont {R.}~\bibnamefont {Fitzpatrick}},\ }\bibfield
  {title} {\enquote {\bibinfo {title} {Role of singular layers in the plasma
  response to resonant magnetic perturbations},}\ }\href@noop {} {\bibfield
  {journal} {\bibinfo  {journal} {Nuclear Fusion}\ }\textbf {\bibinfo {volume}
  {52}},\ \bibinfo {pages} {074004} (\bibinfo {year} {2012})}\BibitemShut
  {NoStop}%
\bibitem [{\citenamefont {Koslowski}\ \emph {et~al.}(2006)\citenamefont
  {Koslowski}, \citenamefont {Liang}, \citenamefont {Kr{\"a}mer-Flecken},
  \citenamefont {L{\"o}wenbr{\"u}ck}, \citenamefont {Von~Hellermann},
  \citenamefont {Westerhof}, \citenamefont {Wolf}, \citenamefont {Zimmermann},
  \citenamefont {team} \emph {et~al.}}]{koslowski2006dependence}%
  \BibitemOpen
  \bibfield  {author} {\bibinfo {author} {\bibfnamefont {H.}~\bibnamefont
  {Koslowski}}, \bibinfo {author} {\bibfnamefont {Y.}~\bibnamefont {Liang}},
  \bibinfo {author} {\bibfnamefont {A.}~\bibnamefont {Kr{\"a}mer-Flecken}},
  \bibinfo {author} {\bibfnamefont {K.}~\bibnamefont {L{\"o}wenbr{\"u}ck}},
  \bibinfo {author} {\bibfnamefont {M.}~\bibnamefont {Von~Hellermann}},
  \bibinfo {author} {\bibfnamefont {E.}~\bibnamefont {Westerhof}}, \bibinfo
  {author} {\bibfnamefont {R.}~\bibnamefont {Wolf}}, \bibinfo {author}
  {\bibfnamefont {O.}~\bibnamefont {Zimmermann}}, \bibinfo {author}
  {\bibfnamefont {T.}~\bibnamefont {team}},  \emph {et~al.},\ }\bibfield
  {title} {\enquote {\bibinfo {title} {Dependence of the threshold for
  perturbation field generated m/n= 2/1 tearing modes on the plasma fluid
  rotation},}\ }\href@noop {} {\bibfield  {journal} {\bibinfo  {journal}
  {Nuclear Fusion}\ }\textbf {\bibinfo {volume} {46}},\ \bibinfo {pages} {L1}
  (\bibinfo {year} {2006})}\BibitemShut {NoStop}%
\bibitem [{\citenamefont {Heyn}\ \emph {et~al.}(2006)\citenamefont {Heyn},
  \citenamefont {Ivanov}, \citenamefont {Kasilov},\ and\ \citenamefont
  {Kernbichler}}]{heyn2006kinetic}%
  \BibitemOpen
  \bibfield  {author} {\bibinfo {author} {\bibfnamefont {M.~F.}\ \bibnamefont
  {Heyn}}, \bibinfo {author} {\bibfnamefont {I.~B.}\ \bibnamefont {Ivanov}},
  \bibinfo {author} {\bibfnamefont {S.~V.}\ \bibnamefont {Kasilov}}, \ and\
  \bibinfo {author} {\bibfnamefont {W.}~\bibnamefont {Kernbichler}},\
  }\bibfield  {title} {\enquote {\bibinfo {title} {Kinetic modelling of the
  interaction of rotating magnetic fields with a radially inhomogeneous
  plasma},}\ }\href@noop {} {\bibfield  {journal} {\bibinfo  {journal} {Nuclear
  fusion}\ }\textbf {\bibinfo {volume} {46}},\ \bibinfo {pages} {S159}
  (\bibinfo {year} {2006})}\BibitemShut {NoStop}%
\bibitem [{\citenamefont {Kikuchi}\ \emph {et~al.}(2006)\citenamefont
  {Kikuchi}, \citenamefont {de~Bock}, \citenamefont {Finken}, \citenamefont
  {Jakubowski}, \citenamefont {Jaspers}, \citenamefont {Koslowski},
  \citenamefont {Kraemer-Flecken}, \citenamefont {Lehnen}, \citenamefont
  {Liang}, \citenamefont {Matsunaga} \emph {et~al.}}]{kikuchi2006forced}%
  \BibitemOpen
  \bibfield  {author} {\bibinfo {author} {\bibfnamefont {Y.}~\bibnamefont
  {Kikuchi}}, \bibinfo {author} {\bibfnamefont {M.}~\bibnamefont {de~Bock}},
  \bibinfo {author} {\bibfnamefont {K.}~\bibnamefont {Finken}}, \bibinfo
  {author} {\bibfnamefont {M.}~\bibnamefont {Jakubowski}}, \bibinfo {author}
  {\bibfnamefont {R.}~\bibnamefont {Jaspers}}, \bibinfo {author} {\bibfnamefont
  {H.}~\bibnamefont {Koslowski}}, \bibinfo {author} {\bibfnamefont
  {A.}~\bibnamefont {Kraemer-Flecken}}, \bibinfo {author} {\bibfnamefont
  {M.}~\bibnamefont {Lehnen}}, \bibinfo {author} {\bibfnamefont
  {Y.}~\bibnamefont {Liang}}, \bibinfo {author} {\bibfnamefont
  {G.}~\bibnamefont {Matsunaga}},  \emph {et~al.},\ }\bibfield  {title}
  {\enquote {\bibinfo {title} {Forced magnetic reconnection and field
  penetration of an externally applied rotating helical magnetic field in the
  {TEXTOR} tokamak},}\ }\href@noop {} {\bibfield  {journal} {\bibinfo
  {journal} {Physical review letters}\ }\textbf {\bibinfo {volume} {97}},\
  \bibinfo {pages} {085003} (\bibinfo {year} {2006})}\BibitemShut {NoStop}%
\bibitem [{\citenamefont {Yu}\ \emph {et~al.}(2008)\citenamefont {Yu},
  \citenamefont {G{\"u}nter}, \citenamefont {Kikuchi},\ and\ \citenamefont
  {Finken}}]{yu2008numerical}%
  \BibitemOpen
  \bibfield  {author} {\bibinfo {author} {\bibfnamefont {Q.}~\bibnamefont
  {Yu}}, \bibinfo {author} {\bibfnamefont {S.}~\bibnamefont {G{\"u}nter}},
  \bibinfo {author} {\bibfnamefont {Y.}~\bibnamefont {Kikuchi}}, \ and\
  \bibinfo {author} {\bibfnamefont {K.}~\bibnamefont {Finken}},\ }\bibfield
  {title} {\enquote {\bibinfo {title} {Numerical modelling of error field
  penetration},}\ }\href@noop {} {\bibfield  {journal} {\bibinfo  {journal}
  {Nuclear fusion}\ }\textbf {\bibinfo {volume} {48}},\ \bibinfo {pages}
  {024007} (\bibinfo {year} {2008})}\BibitemShut {NoStop}%
\bibitem [{\citenamefont {Becoulet}\ \emph {et~al.}(2012)\citenamefont
  {Becoulet}, \citenamefont {Orain}, \citenamefont {Maget}, \citenamefont
  {Mellet}, \citenamefont {Garbet}, \citenamefont {Nardon}, \citenamefont
  {Huysmans}, \citenamefont {Casper}, \citenamefont {Loarte}, \citenamefont
  {Cahyna} \emph {et~al.}}]{becoulet2012screening}%
  \BibitemOpen
  \bibfield  {author} {\bibinfo {author} {\bibfnamefont {M.}~\bibnamefont
  {Becoulet}}, \bibinfo {author} {\bibfnamefont {F.}~\bibnamefont {Orain}},
  \bibinfo {author} {\bibfnamefont {P.}~\bibnamefont {Maget}}, \bibinfo
  {author} {\bibfnamefont {N.}~\bibnamefont {Mellet}}, \bibinfo {author}
  {\bibfnamefont {X.}~\bibnamefont {Garbet}}, \bibinfo {author} {\bibfnamefont
  {E.}~\bibnamefont {Nardon}}, \bibinfo {author} {\bibfnamefont
  {G.}~\bibnamefont {Huysmans}}, \bibinfo {author} {\bibfnamefont
  {T.}~\bibnamefont {Casper}}, \bibinfo {author} {\bibfnamefont
  {A.}~\bibnamefont {Loarte}}, \bibinfo {author} {\bibfnamefont
  {P.}~\bibnamefont {Cahyna}},  \emph {et~al.},\ }\bibfield  {title} {\enquote
  {\bibinfo {title} {Screening of resonant magnetic perturbations by flows in
  tokamaks},}\ }\href@noop {} {\bibfield  {journal} {\bibinfo  {journal}
  {Nuclear Fusion}\ }\textbf {\bibinfo {volume} {52}},\ \bibinfo {pages}
  {054003} (\bibinfo {year} {2012})}\BibitemShut {NoStop}%
\bibitem [{\citenamefont {Moyer}\ \emph {et~al.}(2017)\citenamefont {Moyer},
  \citenamefont {Paz-Soldan}, \citenamefont {Nazikian}, \citenamefont {Orlov},
  \citenamefont {Ferraro}, \citenamefont {Grierson}, \citenamefont
  {Kn{\"o}lker}, \citenamefont {Lyons}, \citenamefont {McKee}, \citenamefont
  {Osborne} \emph {et~al.}}]{moyer2017validation}%
  \BibitemOpen
  \bibfield  {author} {\bibinfo {author} {\bibfnamefont {R.~A.}\ \bibnamefont
  {Moyer}}, \bibinfo {author} {\bibfnamefont {C.}~\bibnamefont {Paz-Soldan}},
  \bibinfo {author} {\bibfnamefont {R.}~\bibnamefont {Nazikian}}, \bibinfo
  {author} {\bibfnamefont {D.~M.}\ \bibnamefont {Orlov}}, \bibinfo {author}
  {\bibfnamefont {N.}~\bibnamefont {Ferraro}}, \bibinfo {author} {\bibfnamefont
  {B.~A.}\ \bibnamefont {Grierson}}, \bibinfo {author} {\bibfnamefont
  {M.}~\bibnamefont {Kn{\"o}lker}}, \bibinfo {author} {\bibfnamefont
  {B.}~\bibnamefont {Lyons}}, \bibinfo {author} {\bibfnamefont {G.~R.}\
  \bibnamefont {McKee}}, \bibinfo {author} {\bibfnamefont {T.~H.}\ \bibnamefont
  {Osborne}},  \emph {et~al.},\ }\bibfield  {title} {\enquote {\bibinfo {title}
  {Validation of the model for {ELM} suppression with {3D} magnetic fields
  using low torque {ITER} baseline scenario discharges in {DIII-D}},}\
  }\href@noop {} {\bibfield  {journal} {\bibinfo  {journal} {Physics of
  Plasmas}\ }\textbf {\bibinfo {volume} {24}},\ \bibinfo {pages} {102501}
  (\bibinfo {year} {2017})}\BibitemShut {NoStop}%
\bibitem [{\citenamefont {Lyons}\ \emph {et~al.}(2017)\citenamefont {Lyons},
  \citenamefont {Ferraro}, \citenamefont {Paz-Soldan}, \citenamefont
  {Nazikian},\ and\ \citenamefont {Wingen}}]{lyons2017effect}%
  \BibitemOpen
  \bibfield  {author} {\bibinfo {author} {\bibfnamefont {B.~C.}\ \bibnamefont
  {Lyons}}, \bibinfo {author} {\bibfnamefont {N.~M.}\ \bibnamefont {Ferraro}},
  \bibinfo {author} {\bibfnamefont {C.}~\bibnamefont {Paz-Soldan}}, \bibinfo
  {author} {\bibfnamefont {R.}~\bibnamefont {Nazikian}}, \ and\ \bibinfo
  {author} {\bibfnamefont {A.}~\bibnamefont {Wingen}},\ }\bibfield  {title}
  {\enquote {\bibinfo {title} {Effect of rotation zero-crossing on single-fluid
  plasma response to three-dimensional magnetic perturbations},}\ }\href@noop
  {} {\bibfield  {journal} {\bibinfo  {journal} {Plasma Physics and Controlled
  Fusion}\ }\textbf {\bibinfo {volume} {59}},\ \bibinfo {pages} {044001}
  (\bibinfo {year} {2017})}\BibitemShut {NoStop}%
\bibitem [{\citenamefont {Paz-Soldan}\ \emph {et~al.}(2019)\citenamefont
  {Paz-Soldan}, \citenamefont {Nazikian}, \citenamefont {Cui}, \citenamefont
  {Lyons}, \citenamefont {Orlov}, \citenamefont {Kirk}, \citenamefont {Logan},
  \citenamefont {Osborne}, \citenamefont {Suttrop},\ and\ \citenamefont
  {Weisberg}}]{paz2019effect}%
  \BibitemOpen
  \bibfield  {author} {\bibinfo {author} {\bibfnamefont {C.}~\bibnamefont
  {Paz-Soldan}}, \bibinfo {author} {\bibfnamefont {R.}~\bibnamefont
  {Nazikian}}, \bibinfo {author} {\bibfnamefont {L.}~\bibnamefont {Cui}},
  \bibinfo {author} {\bibfnamefont {B.}~\bibnamefont {Lyons}}, \bibinfo
  {author} {\bibfnamefont {D.}~\bibnamefont {Orlov}}, \bibinfo {author}
  {\bibfnamefont {A.}~\bibnamefont {Kirk}}, \bibinfo {author} {\bibfnamefont
  {N.}~\bibnamefont {Logan}}, \bibinfo {author} {\bibfnamefont
  {T.}~\bibnamefont {Osborne}}, \bibinfo {author} {\bibfnamefont
  {W.}~\bibnamefont {Suttrop}}, \ and\ \bibinfo {author} {\bibfnamefont
  {D.}~\bibnamefont {Weisberg}},\ }\bibfield  {title} {\enquote {\bibinfo
  {title} {The effect of plasma shape and neutral beam mix on the rotation
  threshold for {RMP-ELM} suppression},}\ }\href@noop {} {\bibfield  {journal}
  {\bibinfo  {journal} {Nuclear Fusion}\ }\textbf {\bibinfo {volume} {59}},\
  \bibinfo {pages} {056012} (\bibinfo {year} {2019})}\BibitemShut {NoStop}%
\bibitem [{\citenamefont {Liu}\ \emph {et~al.}(2000)\citenamefont {Liu},
  \citenamefont {Bondeson}, \citenamefont {Fransson}, \citenamefont
  {Lennartson},\ and\ \citenamefont {Breitholtz}}]{liu2000feedback}%
  \BibitemOpen
  \bibfield  {author} {\bibinfo {author} {\bibfnamefont {Y.}~\bibnamefont
  {Liu}}, \bibinfo {author} {\bibfnamefont {A.}~\bibnamefont {Bondeson}},
  \bibinfo {author} {\bibfnamefont {C.-M.}\ \bibnamefont {Fransson}}, \bibinfo
  {author} {\bibfnamefont {B.}~\bibnamefont {Lennartson}}, \ and\ \bibinfo
  {author} {\bibfnamefont {C.}~\bibnamefont {Breitholtz}},\ }\bibfield  {title}
  {\enquote {\bibinfo {title} {Feedback stabilization of nonaxisymmetric
  resistive wall modes in tokamaks. {I}. electromagnetic model},}\ }\href@noop
  {} {\bibfield  {journal} {\bibinfo  {journal} {Physics of Plasmas}\ }\textbf
  {\bibinfo {volume} {7}},\ \bibinfo {pages} {3681--3690} (\bibinfo {year}
  {2000})}\BibitemShut {NoStop}%
\bibitem [{\citenamefont {Turnbull}\ \emph {et~al.}(2013)\citenamefont
  {Turnbull}, \citenamefont {Ferraro}, \citenamefont {Izzo}, \citenamefont
  {Lazarus}, \citenamefont {Park}, \citenamefont {Cooper}, \citenamefont
  {Hirshman}, \citenamefont {Lao}, \citenamefont {Lanctot}, \citenamefont
  {Lazerson} \emph {et~al.}}]{turnbull2013comparisons}%
  \BibitemOpen
  \bibfield  {author} {\bibinfo {author} {\bibfnamefont {A.}~\bibnamefont
  {Turnbull}}, \bibinfo {author} {\bibfnamefont {N.}~\bibnamefont {Ferraro}},
  \bibinfo {author} {\bibfnamefont {V.}~\bibnamefont {Izzo}}, \bibinfo {author}
  {\bibfnamefont {E.~A.}\ \bibnamefont {Lazarus}}, \bibinfo {author}
  {\bibfnamefont {J.-K.}\ \bibnamefont {Park}}, \bibinfo {author}
  {\bibfnamefont {W.}~\bibnamefont {Cooper}}, \bibinfo {author} {\bibfnamefont
  {S.~P.}\ \bibnamefont {Hirshman}}, \bibinfo {author} {\bibfnamefont {L.~L.}\
  \bibnamefont {Lao}}, \bibinfo {author} {\bibfnamefont {M.}~\bibnamefont
  {Lanctot}}, \bibinfo {author} {\bibfnamefont {S.}~\bibnamefont {Lazerson}},
  \emph {et~al.},\ }\bibfield  {title} {\enquote {\bibinfo {title} {Comparisons
  of linear and nonlinear plasma response models for non-axisymmetric
  perturbations},}\ }\href@noop {} {\bibfield  {journal} {\bibinfo  {journal}
  {Physics of Plasmas}\ }\textbf {\bibinfo {volume} {20}},\ \bibinfo {pages}
  {056114} (\bibinfo {year} {2013})}\BibitemShut {NoStop}%
\bibitem [{\citenamefont {Lanctot}\ \emph {et~al.}(2010)\citenamefont
  {Lanctot}, \citenamefont {Reimerdes}, \citenamefont {Garofalo}, \citenamefont
  {Chu}, \citenamefont {Liu}, \citenamefont {Strait}, \citenamefont {Jackson},
  \citenamefont {La~Haye}, \citenamefont {Okabayashi}, \citenamefont {Osborne}
  \emph {et~al.}}]{lanctot2010validation}%
  \BibitemOpen
  \bibfield  {author} {\bibinfo {author} {\bibfnamefont {M.}~\bibnamefont
  {Lanctot}}, \bibinfo {author} {\bibfnamefont {H.}~\bibnamefont {Reimerdes}},
  \bibinfo {author} {\bibfnamefont {A.}~\bibnamefont {Garofalo}}, \bibinfo
  {author} {\bibfnamefont {M.}~\bibnamefont {Chu}}, \bibinfo {author}
  {\bibfnamefont {Y.}~\bibnamefont {Liu}}, \bibinfo {author} {\bibfnamefont
  {E.}~\bibnamefont {Strait}}, \bibinfo {author} {\bibfnamefont
  {G.}~\bibnamefont {Jackson}}, \bibinfo {author} {\bibfnamefont
  {R.}~\bibnamefont {La~Haye}}, \bibinfo {author} {\bibfnamefont
  {M.}~\bibnamefont {Okabayashi}}, \bibinfo {author} {\bibfnamefont
  {T.}~\bibnamefont {Osborne}},  \emph {et~al.},\ }\bibfield  {title} {\enquote
  {\bibinfo {title} {Validation of the linear ideal magnetohydrodynamic model
  of three-dimensional tokamak equilibria},}\ }\href@noop {} {\bibfield
  {journal} {\bibinfo  {journal} {Physics of Plasmas}\ }\textbf {\bibinfo
  {volume} {17}},\ \bibinfo {pages} {030701} (\bibinfo {year}
  {2010})}\BibitemShut {NoStop}%
\bibitem [{\citenamefont {Wang}\ \emph {et~al.}(2015)\citenamefont {Wang},
  \citenamefont {Lanctot}, \citenamefont {Liu}, \citenamefont {Park},\ and\
  \citenamefont {Menard}}]{wang2015three}%
  \BibitemOpen
  \bibfield  {author} {\bibinfo {author} {\bibfnamefont {Z.~R.}\ \bibnamefont
  {Wang}}, \bibinfo {author} {\bibfnamefont {M.~J.}\ \bibnamefont {Lanctot}},
  \bibinfo {author} {\bibfnamefont {Y.}~\bibnamefont {Liu}}, \bibinfo {author}
  {\bibfnamefont {J.-K.}\ \bibnamefont {Park}}, \ and\ \bibinfo {author}
  {\bibfnamefont {J.~E.}\ \bibnamefont {Menard}},\ }\bibfield  {title}
  {\enquote {\bibinfo {title} {Three-dimensional drift kinetic response of
  high-$\beta$ plasmas in the {DIII-D} tokamak},}\ }\href@noop {} {\bibfield
  {journal} {\bibinfo  {journal} {Physical Review Letters}\ }\textbf {\bibinfo
  {volume} {114}},\ \bibinfo {pages} {145005} (\bibinfo {year}
  {2015})}\BibitemShut {NoStop}%
\bibitem [{\citenamefont {Liu}\ \emph {et~al.}(2011)\citenamefont {Liu},
  \citenamefont {Kirk}, \citenamefont {Gribov}, \citenamefont {Gryaznevich},
  \citenamefont {Hender},\ and\ \citenamefont {Nardon}}]{liu2011modelling}%
  \BibitemOpen
  \bibfield  {author} {\bibinfo {author} {\bibfnamefont {Y.}~\bibnamefont
  {Liu}}, \bibinfo {author} {\bibfnamefont {A.}~\bibnamefont {Kirk}}, \bibinfo
  {author} {\bibfnamefont {Y.}~\bibnamefont {Gribov}}, \bibinfo {author}
  {\bibfnamefont {M.}~\bibnamefont {Gryaznevich}}, \bibinfo {author}
  {\bibfnamefont {T.}~\bibnamefont {Hender}}, \ and\ \bibinfo {author}
  {\bibfnamefont {E.}~\bibnamefont {Nardon}},\ }\bibfield  {title} {\enquote
  {\bibinfo {title} {Modelling of plasma response to resonant magnetic
  perturbation fields in {MAST} and {ITER}},}\ }\href@noop {} {\bibfield
  {journal} {\bibinfo  {journal} {Nuclear Fusion}\ }\textbf {\bibinfo {volume}
  {51}},\ \bibinfo {pages} {083002} (\bibinfo {year} {2011})}\BibitemShut
  {NoStop}%
\bibitem [{\citenamefont {Wan}\ \emph {et~al.}(2019)\citenamefont {Wan},
  \citenamefont {Liang}, \citenamefont {Gong}, \citenamefont {Xiang},
  \citenamefont {Xu}, \citenamefont {Sun}, \citenamefont {Wang}, \citenamefont
  {Qian}, \citenamefont {Liu}, \citenamefont {Zeng} \emph
  {et~al.}}]{wan2019recent}%
  \BibitemOpen
  \bibfield  {author} {\bibinfo {author} {\bibfnamefont {B.}~\bibnamefont
  {Wan}}, \bibinfo {author} {\bibfnamefont {Y.}~\bibnamefont {Liang}}, \bibinfo
  {author} {\bibfnamefont {X.}~\bibnamefont {Gong}}, \bibinfo {author}
  {\bibfnamefont {N.}~\bibnamefont {Xiang}}, \bibinfo {author} {\bibfnamefont
  {G.}~\bibnamefont {Xu}}, \bibinfo {author} {\bibfnamefont {Y.}~\bibnamefont
  {Sun}}, \bibinfo {author} {\bibfnamefont {L.}~\bibnamefont {Wang}}, \bibinfo
  {author} {\bibfnamefont {J.}~\bibnamefont {Qian}}, \bibinfo {author}
  {\bibfnamefont {H.}~\bibnamefont {Liu}}, \bibinfo {author} {\bibfnamefont
  {L.}~\bibnamefont {Zeng}},  \emph {et~al.},\ }\bibfield  {title} {\enquote
  {\bibinfo {title} {Recent advances in east physics experiments in support of
  steady-state operation for {ITER} and {CFETR}},}\ }\href@noop {} {\bibfield
  {journal} {\bibinfo  {journal} {Nuclear Fusion}\ }\textbf {\bibinfo {volume}
  {59}},\ \bibinfo {pages} {112003} (\bibinfo {year} {2019})}\BibitemShut
  {NoStop}%
\bibitem [{\citenamefont {Wan}\ \emph {et~al.}(2020)\citenamefont {Wan} \emph
  {et~al.}}]{wan2020new}%
  \BibitemOpen
  \bibfield  {author} {\bibinfo {author} {\bibfnamefont {B.}~\bibnamefont
  {Wan}} \emph {et~al.},\ }\bibfield  {title} {\enquote {\bibinfo {title} {A
  new path to improve high $\beta$p plasma performance on {EAST} for
  steady-state tokamak fusion reactor},}\ }\href@noop {} {\bibfield  {journal}
  {\bibinfo  {journal} {Chinese Physics Letters}\ }\textbf {\bibinfo {volume}
  {37}},\ \bibinfo {pages} {045202} (\bibinfo {year} {2020})}\BibitemShut
  {NoStop}%
\bibitem [{\citenamefont {Liu}\ \emph {et~al.}(2014)\citenamefont {Liu},
  \citenamefont {Kirk}, \citenamefont {Thornton}, \citenamefont {Team} \emph
  {et~al.}}]{liu2014modelling}%
  \BibitemOpen
  \bibfield  {author} {\bibinfo {author} {\bibfnamefont {Y.}~\bibnamefont
  {Liu}}, \bibinfo {author} {\bibfnamefont {A.}~\bibnamefont {Kirk}}, \bibinfo
  {author} {\bibfnamefont {A.}~\bibnamefont {Thornton}}, \bibinfo {author}
  {\bibfnamefont {M.}~\bibnamefont {Team}},  \emph {et~al.},\ }\bibfield
  {title} {\enquote {\bibinfo {title} {Modelling intrinsic error field
  correction experiments in {MAST}},}\ }\href@noop {} {\bibfield  {journal}
  {\bibinfo  {journal} {Plasma Physics and Controlled Fusion}\ }\textbf
  {\bibinfo {volume} {56}},\ \bibinfo {pages} {104002} (\bibinfo {year}
  {2014})}\BibitemShut {NoStop}%
\bibitem [{\citenamefont {Cole}\ and\ \citenamefont
  {Fitzpatrick}(2006)}]{cole2006drift}%
  \BibitemOpen
  \bibfield  {author} {\bibinfo {author} {\bibfnamefont {A.}~\bibnamefont
  {Cole}}\ and\ \bibinfo {author} {\bibfnamefont {R.}~\bibnamefont
  {Fitzpatrick}},\ }\bibfield  {title} {\enquote {\bibinfo {title}
  {Drift-magnetohydrodynamical model of error-field penetration in tokamak
  plasmas},}\ }\href@noop {} {\bibfield  {journal} {\bibinfo  {journal}
  {Physics of plasmas}\ }\textbf {\bibinfo {volume} {13}},\ \bibinfo {pages}
  {032503} (\bibinfo {year} {2006})}\BibitemShut {NoStop}%
\bibitem [{\citenamefont {Liu}\ \emph {et~al.}(2017)\citenamefont {Liu},
  \citenamefont {Kirk}, \citenamefont {Li}, \citenamefont {In}, \citenamefont
  {Nazikian}, \citenamefont {Sun}, \citenamefont {Suttrop}, \citenamefont
  {Lyons}, \citenamefont {Ryan}, \citenamefont {Wang} \emph
  {et~al.}}]{liu2017comparative}%
  \BibitemOpen
  \bibfield  {author} {\bibinfo {author} {\bibfnamefont {Y.}~\bibnamefont
  {Liu}}, \bibinfo {author} {\bibfnamefont {A.}~\bibnamefont {Kirk}}, \bibinfo
  {author} {\bibfnamefont {L.}~\bibnamefont {Li}}, \bibinfo {author}
  {\bibfnamefont {Y.}~\bibnamefont {In}}, \bibinfo {author} {\bibfnamefont
  {R.}~\bibnamefont {Nazikian}}, \bibinfo {author} {\bibfnamefont
  {Y.}~\bibnamefont {Sun}}, \bibinfo {author} {\bibfnamefont {W.}~\bibnamefont
  {Suttrop}}, \bibinfo {author} {\bibfnamefont {B.}~\bibnamefont {Lyons}},
  \bibinfo {author} {\bibfnamefont {D.}~\bibnamefont {Ryan}}, \bibinfo {author}
  {\bibfnamefont {S.}~\bibnamefont {Wang}},  \emph {et~al.},\ }\bibfield
  {title} {\enquote {\bibinfo {title} {Comparative investigation of {ELM}
  control based on toroidal modelling of plasma response to {RMP} fields},}\
  }\href@noop {} {\bibfield  {journal} {\bibinfo  {journal} {Physics of
  Plasmas}\ }\textbf {\bibinfo {volume} {24}},\ \bibinfo {pages} {056111}
  (\bibinfo {year} {2017})}\BibitemShut {NoStop}%
\bibitem [{\citenamefont {Sun}\ \emph {et~al.}(2015)\citenamefont {Sun},
  \citenamefont {Liang}, \citenamefont {Qian}, \citenamefont {Shen},\ and\
  \citenamefont {Wan}}]{sun2015modeling}%
  \BibitemOpen
  \bibfield  {author} {\bibinfo {author} {\bibfnamefont {Y.}~\bibnamefont
  {Sun}}, \bibinfo {author} {\bibfnamefont {Y.}~\bibnamefont {Liang}}, \bibinfo
  {author} {\bibfnamefont {J.}~\bibnamefont {Qian}}, \bibinfo {author}
  {\bibfnamefont {B.}~\bibnamefont {Shen}}, \ and\ \bibinfo {author}
  {\bibfnamefont {B.}~\bibnamefont {Wan}},\ }\bibfield  {title} {\enquote
  {\bibinfo {title} {Modeling of non-axisymmetric magnetic perturbations in
  tokamaks},}\ }\href@noop {} {\bibfield  {journal} {\bibinfo  {journal}
  {Plasma Physics and Controlled Fusion}\ }\textbf {\bibinfo {volume} {57}},\
  \bibinfo {pages} {045003} (\bibinfo {year} {2015})}\BibitemShut {NoStop}%
\bibitem [{\citenamefont {Liu}\ \emph {et~al.}(2012)\citenamefont {Liu},
  \citenamefont {Connor}, \citenamefont {Cowley}, \citenamefont {Ham},
  \citenamefont {Hastie},\ and\ \citenamefont {Hender}}]{liu2012continuum}%
  \BibitemOpen
  \bibfield  {author} {\bibinfo {author} {\bibfnamefont {Y.}~\bibnamefont
  {Liu}}, \bibinfo {author} {\bibfnamefont {J.}~\bibnamefont {Connor}},
  \bibinfo {author} {\bibfnamefont {S.}~\bibnamefont {Cowley}}, \bibinfo
  {author} {\bibfnamefont {C.}~\bibnamefont {Ham}}, \bibinfo {author}
  {\bibfnamefont {R.}~\bibnamefont {Hastie}}, \ and\ \bibinfo {author}
  {\bibfnamefont {T.}~\bibnamefont {Hender}},\ }\bibfield  {title} {\enquote
  {\bibinfo {title} {Continuum resonance induced electromagnetic torque by a
  rotating plasma response to static resonant magnetic perturbation field},}\
  }\href@noop {} {\bibfield  {journal} {\bibinfo  {journal} {Physics of
  Plasmas}\ }\textbf {\bibinfo {volume} {19}},\ \bibinfo {pages} {102507}
  (\bibinfo {year} {2012})}\BibitemShut {NoStop}%
\bibitem [{\citenamefont {Wesson}\ and\ \citenamefont
  {Campbell}(2011)}]{wesson2011tokamaks}%
  \BibitemOpen
  \bibfield  {author} {\bibinfo {author} {\bibfnamefont {J.}~\bibnamefont
  {Wesson}}\ and\ \bibinfo {author} {\bibfnamefont {D.~J.}\ \bibnamefont
  {Campbell}},\ }\href@noop {} {\emph {\bibinfo {title} {Tokamaks}}},\ Vol.\
  \bibinfo {volume} {149}\ (\bibinfo  {publisher} {Oxford university press},\
  \bibinfo {year} {2011})\BibitemShut {NoStop}%
\bibitem [{\citenamefont {Hahm}\ and\ \citenamefont
  {Burrell}(1995)}]{hahm1995flow}%
  \BibitemOpen
  \bibfield  {author} {\bibinfo {author} {\bibfnamefont {T.}~\bibnamefont
  {Hahm}}\ and\ \bibinfo {author} {\bibfnamefont {K.}~\bibnamefont {Burrell}},\
  }\bibfield  {title} {\enquote {\bibinfo {title} {Flow shear induced
  fluctuation suppression in finite aspect ratio shaped tokamak plasma},}\
  }\href@noop {} {\bibfield  {journal} {\bibinfo  {journal} {Physics of
  Plasmas}\ }\textbf {\bibinfo {volume} {2}},\ \bibinfo {pages} {1648--1651}
  (\bibinfo {year} {1995})}\BibitemShut {NoStop}%
\bibitem [{\citenamefont {Ren}\ \emph {et~al.}(2021)\citenamefont {Ren},
  \citenamefont {Sun}, \citenamefont {Wang}, \citenamefont {Gu}, \citenamefont
  {Qian}, \citenamefont {Shi}, \citenamefont {Shen}, \citenamefont {Liu},
  \citenamefont {Guo}, \citenamefont {Chu} \emph
  {et~al.}}]{ren2021penetration}%
  \BibitemOpen
  \bibfield  {author} {\bibinfo {author} {\bibfnamefont {J.}~\bibnamefont
  {Ren}}, \bibinfo {author} {\bibfnamefont {Y.}~\bibnamefont {Sun}}, \bibinfo
  {author} {\bibfnamefont {H.-H.}\ \bibnamefont {Wang}}, \bibinfo {author}
  {\bibfnamefont {S.}~\bibnamefont {Gu}}, \bibinfo {author} {\bibfnamefont
  {J.}~\bibnamefont {Qian}}, \bibinfo {author} {\bibfnamefont {T.}~\bibnamefont
  {Shi}}, \bibinfo {author} {\bibfnamefont {B.}~\bibnamefont {Shen}}, \bibinfo
  {author} {\bibfnamefont {Y.}~\bibnamefont {Liu}}, \bibinfo {author}
  {\bibfnamefont {W.}~\bibnamefont {Guo}}, \bibinfo {author} {\bibfnamefont
  {N.}~\bibnamefont {Chu}},  \emph {et~al.},\ }\bibfield  {title} {\enquote
  {\bibinfo {title} {Penetration of n= 2 resonant magnetic fieldperturbations
  in {EAST}},}\ }\href@noop {} {\bibfield  {journal} {\bibinfo  {journal}
  {Nuclear Fusion}\ } (\bibinfo {year} {2021})}\BibitemShut {NoStop}%
\end{thebibliography}%

\end{document}